 \documentclass[10pt,pra,aps,a4paper,oneside,twocolumn,superscriptaddress,showpacs]{revtex4-1}

\usepackage[english]{babel}
\usepackage[latin1]{inputenc}
\usepackage[T1]{fontenc}
\usepackage{lmodern}
\usepackage{ulem}
\usepackage{dcolumn}
\usepackage{subfigure}
\usepackage{footnote}
\usepackage{multirow}
\usepackage{amsmath,amsfonts,amssymb}

\usepackage{graphicx}  
\usepackage{latexsym}   
\usepackage{color}

\begin{document}

\title{Tunable Fano resonances in the decay rates of a pointlike emitter near a graphene-coated nanowire}

\author{\firstname{Tiago} J. \surname{Arruda}}
\email{tiagojarruda@gmail.com}
\affiliation{Instituto de F\'isica de S\~ao Carlos,
Universidade de S\~ao Paulo, 13566-590 S\~ao Carlos, S\~ao Paulo, Brazil}

\author{\firstname{Romain} \surname{Bachelard}}
\affiliation{Departamento de F\'isica,
Universidade Federal de S\~ao Carlos, 13565-905 S\~ao Carlos, S\~ao Paulo, Brazil}

\author{\firstname{John} \surname{Weiner}}
\affiliation{Instituto de F\'isica de S\~ao Carlos,
Universidade de S\~ao Paulo, 13566-590 S\~ao Carlos, S\~ao Paulo, Brazil}

\author{\firstname{Philippe} W. \surname{Courteille}}
\affiliation{Instituto de F\'isica de S\~ao Carlos,
Universidade de S\~ao Paulo, 13566-590 S\~ao Carlos, S\~ao Paulo, Brazil}

\begin{abstract}

    Based on the Lorenz-Mie theory, we derive analytical expressions of radiative and nonradiative transition rates for different orientations of a point dipole emitter in the vicinity of an infinitely long circular cylinder of arbitrary radius.
    Special attention is devoted to the spontaneous decay rate of a dipole emitter near a subwavelength-diameter nanowire coated with a graphene monolayer.
    We show that plasmonic Fano resonances associated with light scattering by graphene-coated nanowires appear in the Purcell factor as a function of transition wavelength.
    Furthermore, the Fano line shape of transition rates can be tailored and electrically tuned by varying the distance between emitter and cylinder and by modulating the graphene chemical potential, where the Fano asymmetry parameter is proportional to the square root of the chemical potential.
   This gate-voltage-tunable Fano resonance leads to a resonant enhancement and suppression of light emission in the far-infrared range of frequencies.
   This result could be explored in applications involving ultrahigh-contrast switching for spontaneous emission in specifically designed tunable plasmonic nanostructures.

\end{abstract}


\maketitle


\section{Introduction}

The enhancement or suppression of the spontaneous-emission rate of a quantum emitter induced by its interaction with environment is generally referred to as the Purcell effect~\cite{Purcell_PhysRev69_1946,Chew_JCPhys87_1987,Dereux_PhysRevB84_2011,Bordo_JOSAB31_2014,Belov_SciRep5_2015}.
First described in the context of cavity quantum electrodynamics~\cite{Purcell_PhysRev69_1946}, the Purcell effect finds applications where controlling and manipulating light emission and absorption in subwavelength structures is crucial~\cite{Carminati_SurfSciRep70_2015}, such as single-molecule optical microscopy~\cite{Sandoghdar_Nature405_2000}, high efficiency single-photon sources~\cite{Lukin_NatPhys3_2007,Gallego_PhysRevLett121_2018}, integrated plasmonic amplifiers~\cite{Berini_PhysRevB78_2008}, microcavity light-emitting devices~\cite{Fainman_OptExp21_2013}, and so on.
In recent years, there has been a growing interest in manipulating light emission using nanostructured plasmonic metamaterials~\cite{Farina_PhysRevA87_2013,Liu_Nature9_2014,Belov_SciRep5_2015,Szilard_PhysRevB94_2016,Cuevas_JOpt18_2016,Girard_JOpt18_2016,Arruda_PhysRevA96_2017}.
Among the possibilities to tailor light-matter interaction at ultrasmall lengths, metallic nanostructures have been widely explored to concentrate light at subwavelength scales, owing to the excitation of surface plasmons on metal-insulator interfaces~\cite{Lukin_PhysRevLett97_2006,Lukin_Nature450_2007,Sun_SciApp4_2015,Girard_JOpt18_2016,Gu_SciRep8_2018}.
For perfectly plane surfaces, surface plasmons are nonradiative trapped modes that cannot be excited directly by incident plane waves of infinite extent.
However, for roughened or grooved surfaces, the surface plasmon modes can be coherently excited due to their coupling with incident photons.
Interestingly enough, a dipole emitter in the vicinity of a plasmonic surface can effectively couple photons to surface plasmons even for perfectly plane surfaces~\cite{Philpott_JChemPhys62_1975,Eagen_OptLett4_1979}.
In bounded geometries, these trapped modes are said to be localized and are excited at discrete frequencies that depend on the geometry of the system~\cite{Bohren_Book_1983}.

Recently, graphene has become a promising alternative material to enhance the Purcell factor in subwavelength structures due to its unique optical properties, such as strong localized surface plasmon resonances with relatively lower losses than noble metals~\cite{Soljacic_PhysRevB80_2009,Yakovlev_PhysRevB86_2012}.
Indeed, due to the finite skin depths of metals at infrared and optical frequencies and specific bulk volumetric properties of metallic metamaterials, applications using metallic nanostructures are generally limited by high ohmic losses~\cite{Alu_PhysRevB80_2009,Alu_ACSNano5_2011}.
Conversely, graphene exhibits strong light-matter interaction in two-dimensional atomically thin layers of carbon atoms~\cite{Abajo_NanoLett11_2011}, and it also offers magnetic, electrical, or chemical tunability of its conductivity from terahertz up to midinfrared frequencies~\cite{Wang_NatNano6_2011,Engheta_Sci332_2011,Basov_Nature487_2012,Cuevas_JOpt17_2015,Farina_PhysRevB92_2015,Cuevas_JQSRT173_2016}.

Here, based on the full-wave Lorenz-Mie theory of circular cylinders~\cite{Bohren_Book_1983,Arruda_JOpSocAmA31_2014}, we analytically study the spontaneous-emission rate of a point dipole emitter (e.g., an excited atom, a fluorescent molecule, a quantum dot, or a rare-earth ion) near an infinitely long circular cylinder with dispersive parameters.
Special attention is paid to the Purcell factor of a pointlike emitter in the vicinity of a graphene-coated nanowire.
By varying the distance between emitter and cylinder, the radiative decay rate associated with the dipole moment oriented orthogonal to the cylinder axis is shown to exhibit a Fano line shape as a function of the emission frequency; conversely, a dipole moment oriented parallel to the cylinder axis exhibits a Lorentzian line shape.

First explained in the realm of atomic physics by U. Fano~\cite{Fano_PhysRev124_1961}, the Fano effect has become an important tool for controlling electromagnetic mode interactions and light propagation at a subwavelength scale, owing to advances in micro- and nanofabrication techniques~\cite{Kivshar_RevModPhys82_2010,Wu_SciRep7_2017,Weis_Sci330_2010,Painter_Nature472_2011,Zhou_NatPhys9_2013,Zhou_NatPhys9_2013,Lu_PhysRevApp10_2018,Arruda_Springer219_2018}.
For graphene-coated nanowires, the appearance of a Fano line shape is a signature of interference between a localized plasmon resonance at the graphene coating and a broad dipole resonance acting as a radiation background~\cite{Kivshar_RevModPhys82_2010,Arruda_PhysRevA87_2013,Arruda_PhysRevA96_2017}.
Recent studies have already pointed out the electrical tunability of the Purcell factor using graphene-based nanostructures~\cite{Vidal_PhysRevB85_2012,Jiang_OptExp22_2014,Cuevas_JOpt18_2016} and have shown the presence of asymmetric line shapes in the spectra~\cite{Cuevas_JOpt17_2015,Cuevas_JQSRT200_2017}.
However, the explicit analytical connection between Fano resonances in the Lorenz-Mie theory of cylindrical core-shell scatterers~\cite{Arruda_JOpSocAmA31_2014,Cuevas_JOpt17_2015,Naserpour_SciRep7_2017} and radiative decay rates in the vicinity of graphene-coated nanofibers~\cite{Cuevas_JQSRT200_2017} is still to be established.

In this paper, we formally establish the connection between Fano resonances in light scattering by graphene-coated nanowires and the corresponding Purcell factor of a pointlike emitter.
We demonstrate that both the strong enhancement and suppression of radiative decay rates are associated with plasmonic Lorenz-Mie resonances, where the Fano asymmetry parameter is proportional to the graphene chemical potential.
This allows one to control enhancement and suppression of spontaneous emission by a gate voltage.
In addition, due to their dependence on the Lorenz-Mie coefficients, the analytical expressions of decay rates can be straightforwardly generalized to multilayered cylinders~\cite{Kleiman_JQSRT63_1999}, and they are applicable to any range of frequencies, size parameters, and refractive indices.
These analytical expressions are important to benchmark new numerical tools, such as finite-difference time-domain methods (FDTD)~\cite{Yariv_JOSAB16_1999}, which in turn can be used to characterize more complex geometries.

This paper is organized as follows.
In Sec.~\ref{decay}, we present the theory regarding the decay rates of a pointlike dipole emitter near an arbitrary cylinder, whose basic functions for a core-shell geometry are provided in Appendices~\ref{Lorenz-Mie} and \ref{decay-rates}.
Section~\ref{Graphene} is completely devoted to a graphene-coated nanowire.
In Sec.~\ref{Mie}, we briefly review the light scattering by graphene-coated nanowires under oblique incidence of plane waves and calculate approximate expressions for the scattering efficiencies.
The study of a pointlike dipole source in the vicinity of a graphene-coated nanowire is presented in Sec.~\ref{Fano}, where we show the tunability of Fano resonances in the Purcell factor via a gate voltage.
Finally, in Sec.~\ref{conclusion}, we summarize our main results and conclude.

\section{Decay rates of a dipole emitter in the vicinity of a circular cylinder}
\label{decay}

Within the quantum-mechanical approach, the spontaneous emission of a two-level system is well described by the Fermi golden rule, in which an emitter in the excited state $|{\rm e}\rangle$ decays exponentially to the ground state $|{\rm g}\rangle$.
The corresponding radiative decay rate of this transition can be written as~\cite{Chew_JCPhys87_1987,Klimov_PhysRevA69_2004}
\begin{equation}
\Gamma_{\mathbf{d}_0}^{\rm rad}\propto|\mathbf{d}_0\cdot\mathbf{E}(\mathbf{r}')|^2\rho_{\rm F}(\omega),\label{Gamma-rad-def}
\end{equation}
where $\mathbf{d}_0$ is a nonvanishing matrix element of the dipole moment operator coupling $|{\rm e}\rangle$ to $|{\rm g}\rangle$ and $\mathbf{E}(\mathbf{r}')$ is the electric field amplitude of emitted photon at the emitter position $\mathbf{r}'$ with energy $\hbar\omega$.
The quantity $\rho_{\rm F}(\omega)$ is the final density of photon states (DOS), which is independent of boundary conditions and characterizes the spectral density of eigenmodes of the medium as a whole~\cite{Carminati_SurfSciRep70_2015}.
Usually, one defines the quantity $\rho_{\mathbf{d}_0}(\omega,\mathbf{r}')\equiv|\mathbf{d}_0\cdot\mathbf{E}(\mathbf{r}')|^2\rho_{\rm F}(\omega)/|\mathbf{d}_0|^2$, which is referred to as the local density of states (LDOS) and depends on boundary conditions explicitly~\cite{Carminati_SurfSciRep70_2015}.
In free space, $\mathbf{E}(\mathbf{r}')=\mathbf{E}_{\rm vac}(\mathbf{r}')$ is a plane wave and $\Gamma_{\mathbf{d}_0}^{\rm rad}=\Gamma_0$ is simply the Einstein $A$ coefficient for a quantum emitter: $\Gamma_0\equiv\omega^3|\mathbf{d}_0|^2/3\pi\varepsilon_0\hbar c^3$.

\begin{figure}[htbp]
    \includegraphics[width=\columnwidth]{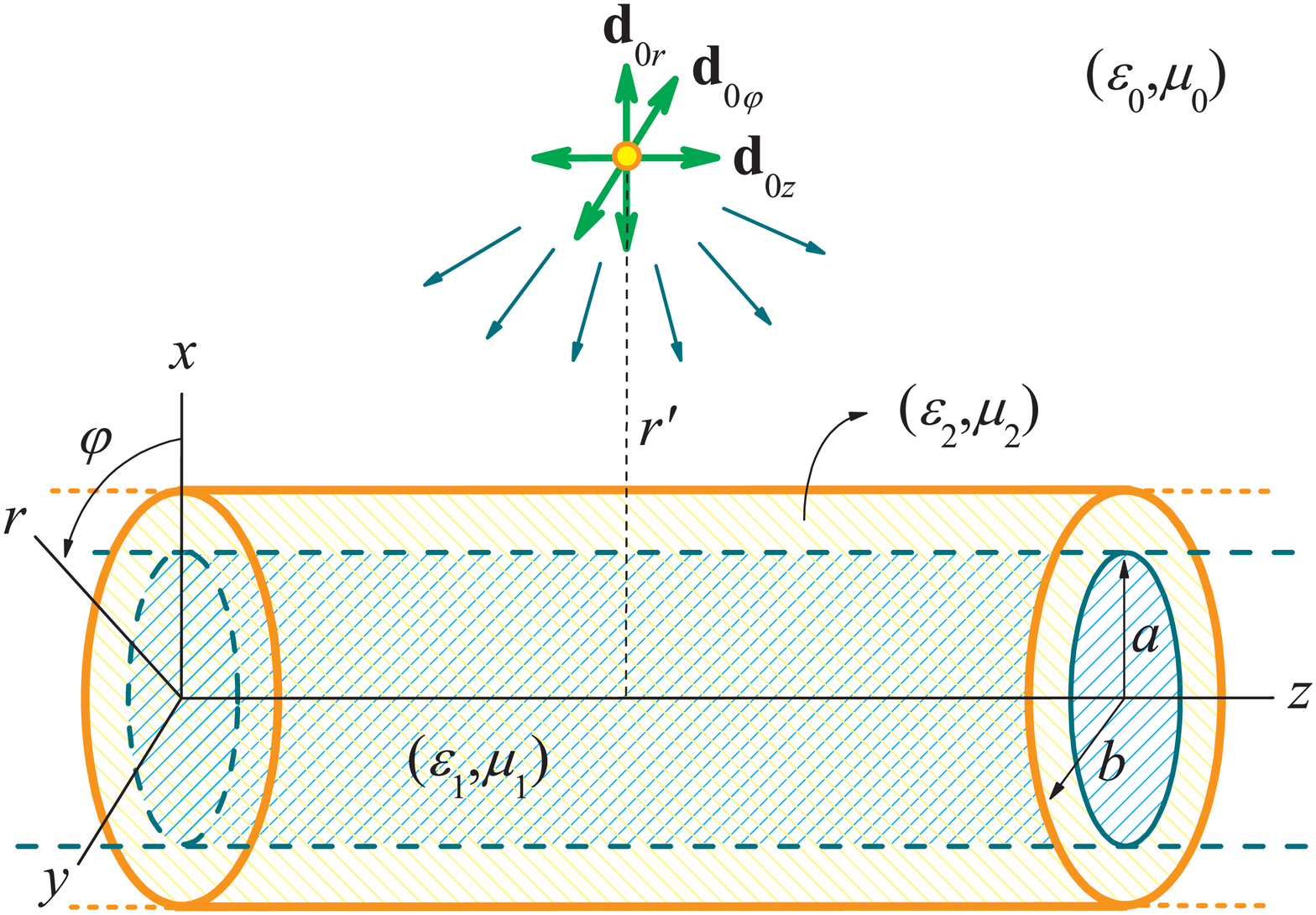}
    \caption{A point dipole emitter in the vicinity of a core-shell cylinder with inner radius $a$ and outer radius $b$.
    The cylinder has optical properties $(\varepsilon_1,\mu_1)$ for the core $(0<r\leq a)$ and $(\varepsilon_2,\mu_2)$ for the shell $(a\leq r\leq b)$, where $\varepsilon$ and $\mu$ are the scalar permittivity and permeability, respectively.
    The surrounding medium is the vacuum ($\varepsilon_0,\mu_0)$.
    There are three basic orientations for the electric dipole moment $\mathbf{d}_0$ in relation to the cylinder: two tangential ($\mathbf{d}_{0z}=d_0\hat{\mathbf{z}}$ and $\mathbf{d}_{0\varphi}=d_0\hat{\boldsymbol{\varphi}}$) and one orthogonal $(\mathbf{d}_{0r}=d_0\hat{\mathbf{r}})$ to the cylindrical surface.}\label{fig1}
\end{figure}

In Eq.~(\ref{Gamma-rad-def}),  $\mathbf{E}(\mathbf{r}')$ is the solution of Maxwell's equations with boundary conditions.
In the presence of a scattering body, one has
\begin{align}
\mathbf{E}(\mathbf{r}')=\mathbf{E}_{\rm vac}(\mathbf{r}')+\mathbf{E}_{\rm sca}(\mathbf{r}'),
\end{align}
where $\mathbf{E}_{\rm sca}(\mathbf{r}')$ is the (returning) field scattered by the scattering body at the position of the emitter.
From now on, let us consider a two-level system located in the vicinity of an infinitely long cylindrical body, as depicted in Fig.~\ref{fig1}.
The scattering plane includes the $z$ axis of the cylinder~\cite{Bohren_Book_1983}.
In free space, the electric fields for transverse-magnetic (TM, $\mathbf{H}_{\rm vac}\perp\hat{\mathbf{z}}$) and transverse-electric (TE, $\mathbf{E}_{\rm vac}\perp\hat{\mathbf{z}}$) modes expanded in cylindrical harmonics for $r\leq r'$ are~\cite{Bohren_Book_1983}:
\begin{subequations}
\begin{align}
\mathbf{E}_{\rm vac}^{\rm TM}(\mathbf{r})&=E_0\left(\sin\zeta\hat{\mathbf{z}}-\cos\zeta\hat{\mathbf{x}}\right) e^{-\imath k(r\sin\zeta\cos\varphi + z\cos\zeta)}\nonumber\\
&=\sum_{\ell=-\infty}^{\infty}E_{\ell}\bigg[
\sin\zeta J_{\ell}(kr\sin\zeta)\hat{\mathbf{z}}\nonumber\\
&\ -\imath\cos\zeta J_{\ell}'(kr\sin\zeta)\hat{\mathbf{r}}+\ell\cos\zeta\frac{J_{\ell}(kr\sin\zeta)}{kr\sin\zeta}\hat{\boldsymbol{\varphi}}\bigg],\label{Einc-TM}\\
\mathbf{E}_{\rm vac}^{\rm TE}(\mathbf{r})&=E_0\hat{\mathbf{y}}e^{-\imath k(r\sin\zeta\cos\varphi + z\cos\zeta)}\nonumber\\
&=\sum_{\ell=-\infty}^{\infty}E_{\ell}\bigg[
\ell\frac{J_{\ell}(kr\sin\zeta)}{kr\sin\zeta}\hat{\mathbf{r}} +\imath J_{\ell}'(kr\sin\zeta)\hat{\boldsymbol{\varphi}}\bigg],\label{Einc-TE}
\end{align}
\end{subequations}
where $E_{\ell}=E_0(-\imath)^{\ell}e^{\imath(\ell\varphi-kz\cos\zeta)}$, $\zeta$ is the angle between the wavevector and the cylinder axis ($0<\zeta<\pi$), and $J_{\ell}(\rho)$ is the cylindrical Bessel function.
For clarity, throughout this paper we omit the time-harmonic dependence $e^{-\imath\omega t}$, where $\omega=k/\sqrt{\varepsilon_0\mu_0}$ is the angular frequency and $\imath^2=-1$.

The field scattered (reflected) by an infinitely long cylinder of radius $b$ is obtained by solving the vector Helmholtz equation in cylindrical coordinates, and then using the continuity of the tangential components of the electromagnetic field.
For the TM polarization, the scattered field in terms of vector cylindrical harmonics for an arbitrary $\zeta$ is~\cite{Bohren_Book_1983}
\begin{align}
&\mathbf{E}_{\rm sca}^{\rm TM}(\mathbf{r})=\sum_{\ell=-\infty}^{\infty}E_{\ell}\bigg\{-b_{\ell}^{\rm TM}\sin\zeta H_{\ell}^{(1)}(kr\sin\zeta)\hat{\mathbf{z}}\nonumber\\
&\ +\left[a_{\ell}^{\rm TM}\ell\frac{H_{\ell}^{(1)}(kr\sin\zeta)}{kr\sin\zeta}+\imath b_{\ell}^{\rm TM}\cos\zeta H_{\ell}'^{(1)}(kr\sin\zeta)\right]\hat{\mathbf{r}}\nonumber\\
&\ +\left[\imath a_{\ell}^{\rm TM} H_{\ell}'^{(1)}(kr\sin\zeta)-b_{\ell}^{\rm TM}\ell\cos\zeta \frac{H_{\ell}^{(1)}(kr\sin\zeta)}{kr\sin\zeta}\right]\hat{\boldsymbol{\varphi}}\bigg\},\label{Esca-TM}
\end{align}
where $H_{\ell}^{(1)}(\rho)$ is the cylindrical Hankel function of the first kind.
The TE polarization is obtained from Eq.~(\ref{Esca-TM}) by replacing the coefficients $(a_{\ell}^{\rm TM},b_{\ell}^{\rm TM})$ with $(-a_{\ell}^{\rm TE},-b_{\ell}^{\rm TE})$.
The Lorenz-Mie coefficients $a_{\ell}^{\rm TM}$, $a_{\ell}^{\rm TE}$, $b_{\ell}^{\rm TM}$, and $b_{\ell}^{\rm TE}$ are explicitly given in Appendix~\ref{Lorenz-Mie} for a core-shell cylinder~\cite{Arruda_JOpSocAmA31_2014}.

In the classical-electrodynamics approach, which provides the same result as the  Weisskopf-Wigner theory, the radiative decay rate is associated with the power radiated by an oscillating dipole normalized to free space~\cite{Chew_JCPhys87_1987}.
Since the power is proportional to $|\mathbf{d}_0\cdot\mathbf{E}(\mathbf{r}')|^2$, where $\mathbf{E}(\mathbf{r}')=\mathbf{E}^{\rm TE}(\mathbf{r}')+\mathbf{E}^{\rm TM}(\mathbf{r}')$ is the local electric field at the emitter position, we have~\cite{Gaponenko_JPhysChem116_2012,Arruda_PhysRevA96_2017}
\begin{align}
\frac{\Gamma_{\mathbf{d}_0}^{\rm rad}(r',\omega)}{\Gamma_{0}}=\frac{\left\langle\left|\mathbf{d}_0\cdot\left[\mathbf{E}_{\rm vac}(\mathbf{r}',\omega)+\mathbf{E}_{\rm sca}(\mathbf{r}',\omega)\right]\right|^2\right\rangle_{\Omega}}{\left\langle\left|\mathbf{d}_0\cdot\mathbf{E}_{\rm vac}(\mathbf{r}',\omega)\right|^2\right\rangle_{\Omega}},\label{definition}
\end{align}
where $\langle \cdots \rangle_{\Omega} =
 (1/4\pi)\int_0^{\pi}{\rm d}\zeta\sin\zeta\int_0^{2\pi}{\rm d}\varphi(\cdots)$ is the angle average over the incidence (polar) angle $\zeta$ and the azimuthal angle $\varphi$, with $\mathbf{d}_0$ being the electric dipole moment in the direction $\hat{\mathbf{d}}_0$.
Note that the same result is obtained by using Eq.~(\ref{Gamma-rad-def}).
Indeed, on the one hand, the angle average comes from the total radiated power calculated by integrating the radial component of the Poynting vector $\mathbf{S}(\mathbf{r})$ in the far field~\cite{Chew_JCPhys87_1987}: $P=\lim_{r\to\infty}r^2\int_{4\pi}{\rm d}\Omega \mathbf{S}(\mathbf{r})\cdot\hat{\mathbf{r}}$.
On the other hand, in the quantum-mechanical approach, the angle average comes from the integration over $k$-space into which the quantum dipole emitter is emitting, where $\rho_{\rm F}(\omega)\propto\delta(\omega_{\mathbf{k}}-\omega)$~\cite{Wylie_PhysRevA30_1984,Farina_PhysRevA87_2013,Carminati_SurfSciRep70_2015}.
For the sake of brevity, the explicit calculation of radiative and nonradiative decay rates within the framework of the Lorenz-Mie theory is provided in Appendix~\ref{decay-rates}.

Throughout this paper, we have used the same notation as Refs.~\cite{Bohren_Book_1983,Arruda_JOpSocAmA31_2014}, which is commonly used for light scattering by an infinite cylinder, where the longitudinal wave number in relation to the fiber is $h=-k\cos\zeta$.
For this reason, instead of integrating over the longitudinal (complex) wave number $h$, we integrate over $\cos\zeta$.
Nonetheless, these two approaches considering real or complex $h$ are equivalent and have the same  fiber eigenvalue equation (poles)~\cite{Abujetas_ACSPhot2_2015}, where the guided mode contribution is derived from the residue~\cite{Klimov_PhysRevA69_2004,
Brandes_PhysRevA79_2009,Girard_JOpt18_2016}.
Here, we emphasize that we are not interested in the separate calculation of guided mode contribution, whose distinction from nonradiative contribution is not well defined for lossy optical fibers~\cite{Klimov_PhysRevA69_2004}.
Moreover, for graphene waveguides, the decay rate near the interface through surface plasmons is shown to be much larger (by over five orders of magnitude) than the decay rate through guided modes~\cite{Cuevas_JOpt18_2016}.

\section{Dielectric nanowire coated with a graphene monolayer}
\label{Graphene}

Let us consider a uniform graphene-coated dielectric nanowire in free space.
We assume a dielectric core made of a lossless material with permittivity $\varepsilon_1\equiv\varepsilon_{\rm d}=3.9\varepsilon_0$ and radius $a=100$~nm.
Since the graphene monolayer is a two-dimensional electromagnetic material and its thickness is much smaller than the radius of the dielectric core, we can treat it as a conducting film~\cite{Alu_ACSNano5_2011,Chen_OptLett40_2015,Cuevas_JOpt17_2015,Naserpour_SciRep7_2017}.
Hence, the graphene monolayer conductivity within the coating nanoshell is well described by using Kubo's formula~\cite{Naserpour_SciRep7_2017,Falkovsky_PhysUSP51_2008}: $\sigma=\sigma_{\rm intra}+\sigma_{\rm inter}$, with the intraband and interband contributions being
\begin{subequations}
\begin{align}
\sigma_{\rm intra} &= \frac{2\imath e^2 k_{\rm B} T}{\pi\hbar^2\left(\omega+\imath\gamma\right)}\ln\left[2\cosh\left(\frac{\mu_{\rm c}}{2k_{\rm B}T}\right)\right],\label{sigma-intra}\\
\sigma_{\rm inter} &= \frac{e^2}{4\hbar}\bigg\{\frac{1}{2}+\frac{1}{\pi}\arctan\left(\frac{\hbar\omega-2\mu_{\rm c}}{2k_{\rm B}T}\right)\nonumber\\
&-\frac{\imath}{2\pi}\ln\left[\frac{(\hbar\omega+2\mu_{\rm c})^2}{(\hbar\omega-2\mu_{\rm c})^2+(2k_{\rm B}T)^2}\right]\bigg\},\label{sigma-inter}
\end{align}
\end{subequations}
where $-e$ is the electron charge, $\hbar$ is the reduced Planck's constant, $k_{\rm B}$ is the Boltzmann's constant, $T$ is the temperature, $\gamma$ is the charge carriers scattering rate, and $\mu_{\rm c}$ is the chemical potential.
By introducing the finite width of the graphene monolayer $t_{\rm g}$, one has a corresponding graphene permittivity $\varepsilon_{\rm g}=\varepsilon_{\rm g}(\omega,\mu_{\rm c},T)$, which is calculated by~\cite{Naserpour_SciRep7_2017}
\begin{align}
\frac{\varepsilon_{\rm g}(\omega)}{\varepsilon_0} = \imath\frac{\sigma(\omega)}{\varepsilon_0\omega t_{\rm g}}.
\end{align}
Here, we consider a nanofiber with an infinite length, i.e., the finite length $L$ of the core-shell cylinder is such that $L\gg\lambda$ and $L\gg b>a$, where $\lambda$ is the operation wavelength and $b$ is the outer radius.
In practice, the fabrication of graphene-coated nanowires is realized in the platform of fiber optics, where a graphene monolayer is wrapped around a single-mode nanofiber.
This nanofiber can be, e.g., a section with the ends tapered down from a standard telecom optical fiber~\cite{Shen_NanoLett14_2014}.

\subsection{Light scattering by a graphene-coated nanowire}
\label{Mie}

\begin{figure*}[htbp!]
\includegraphics[width=.66\columnwidth]{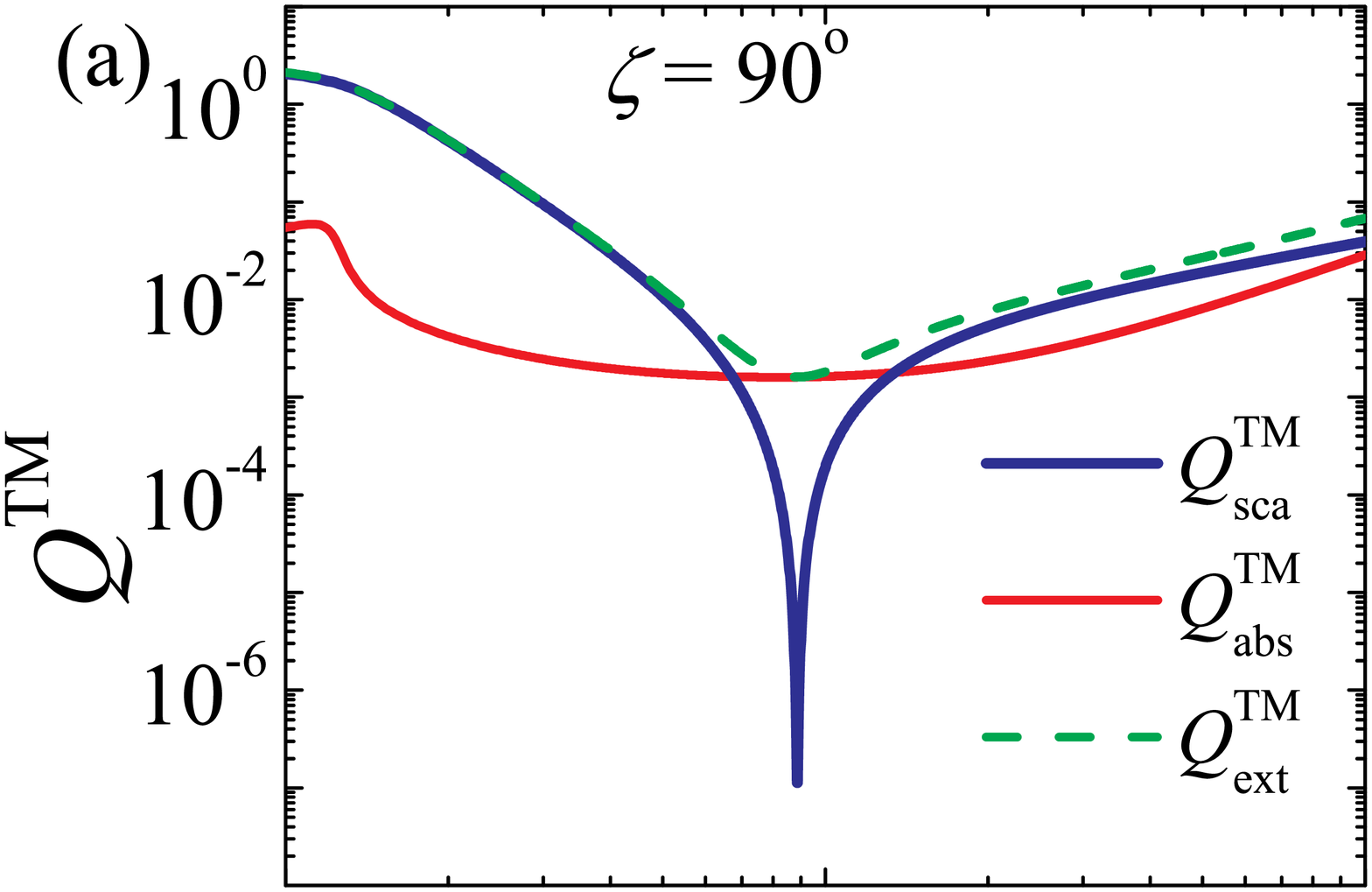}\vspace{-1.2cm}
\includegraphics[width=.66\columnwidth]{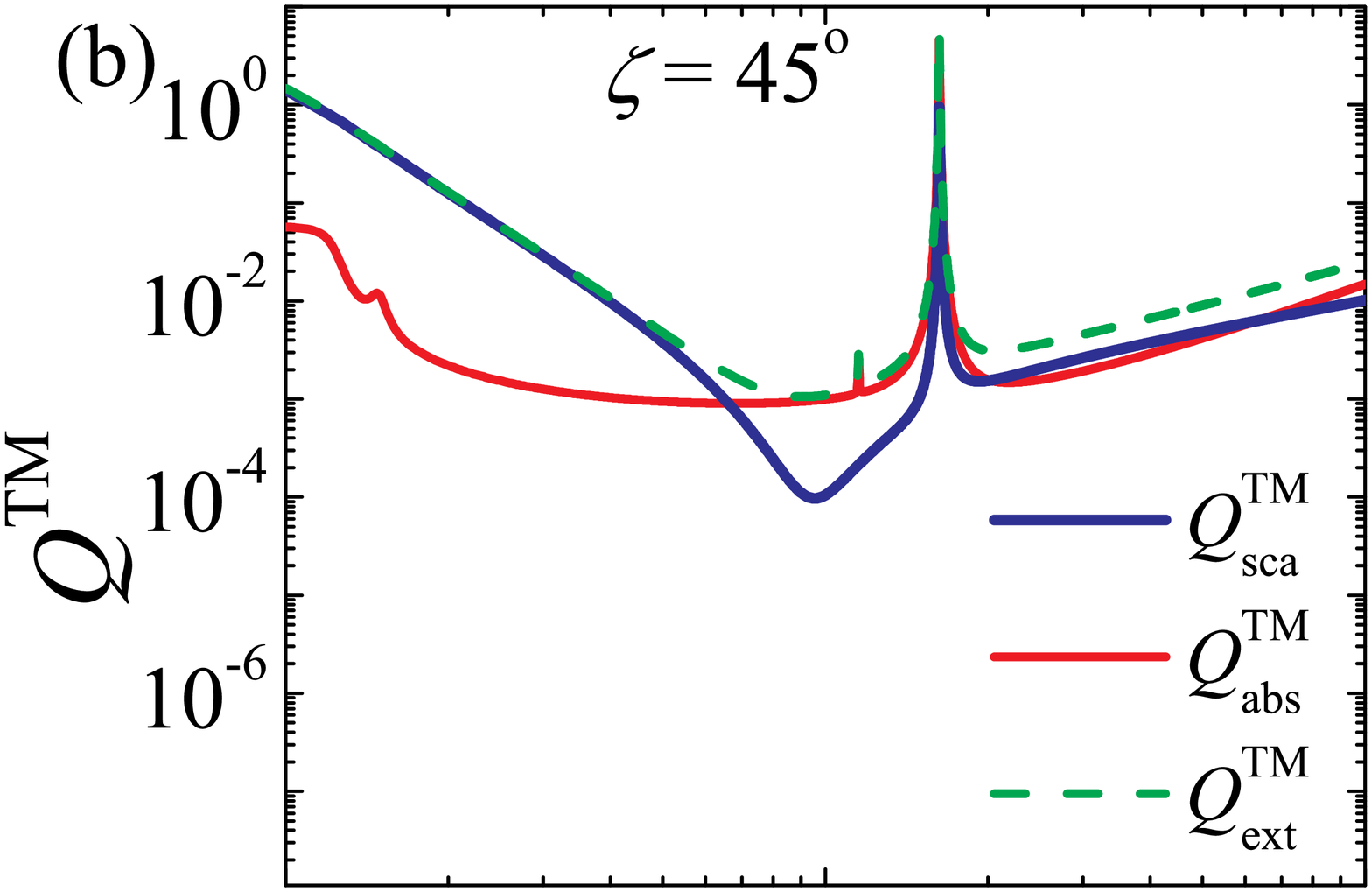}
\includegraphics[width=.66\columnwidth]{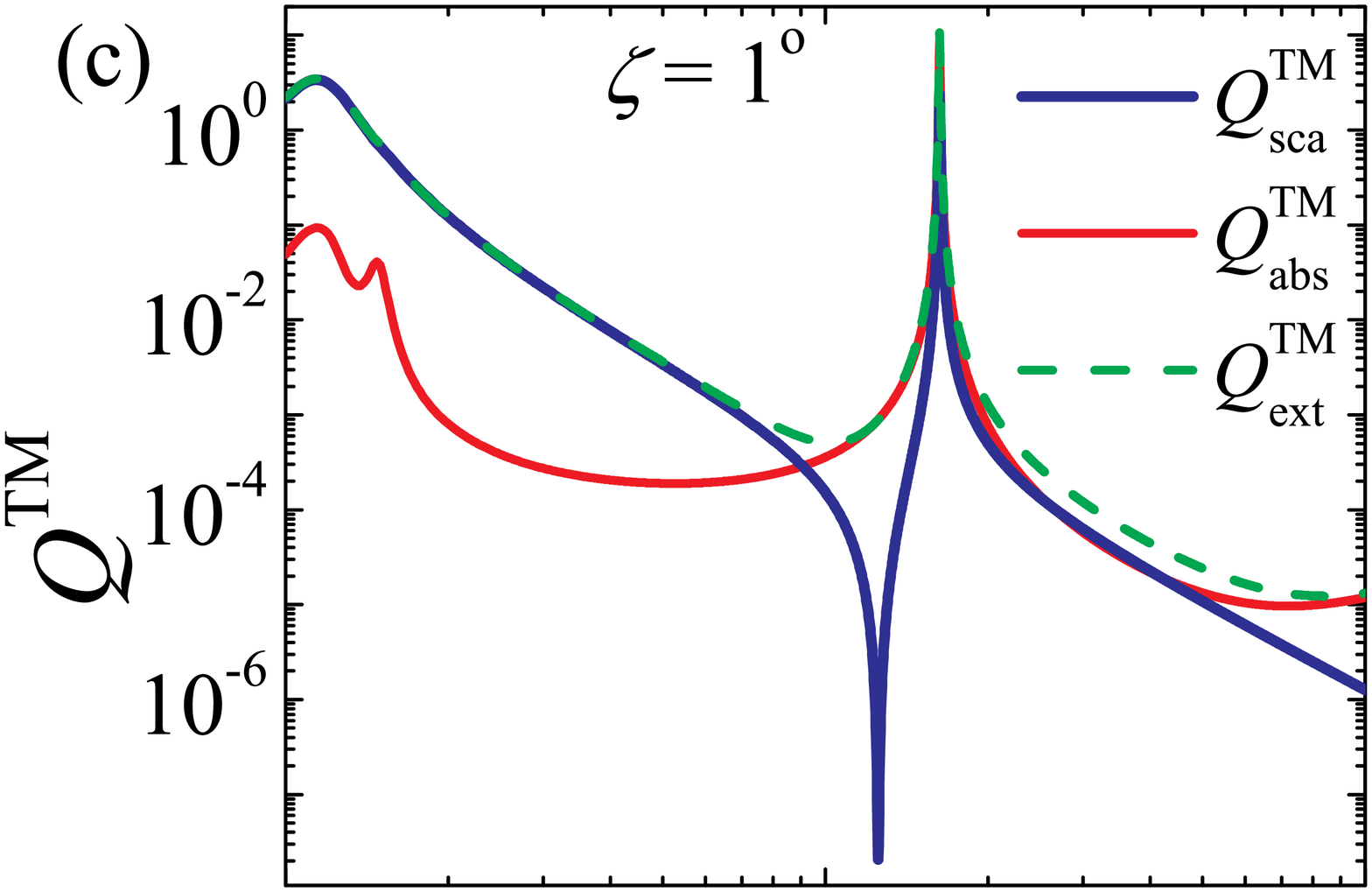}
\includegraphics[width=.66\columnwidth]{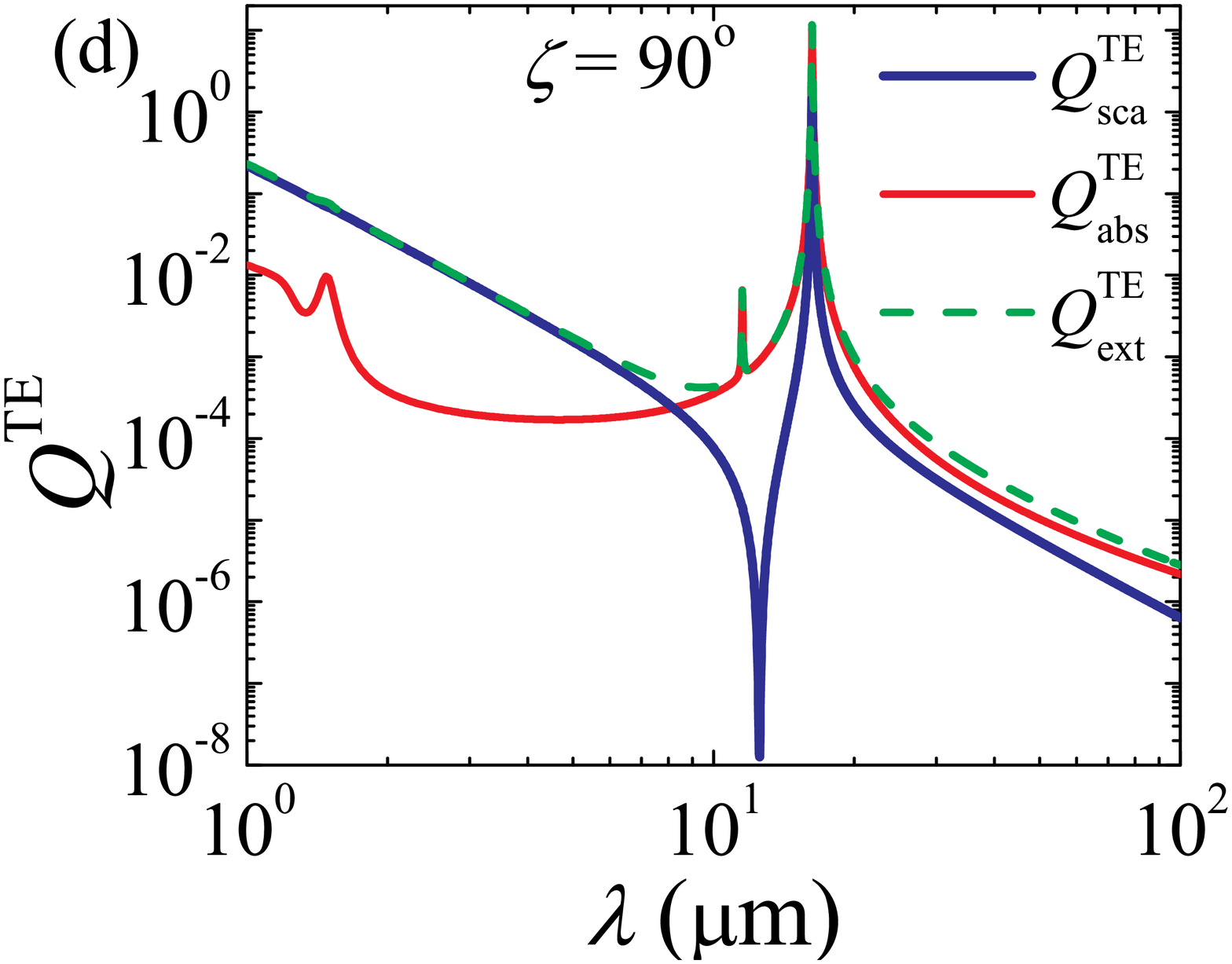}
\includegraphics[width=.66\columnwidth]{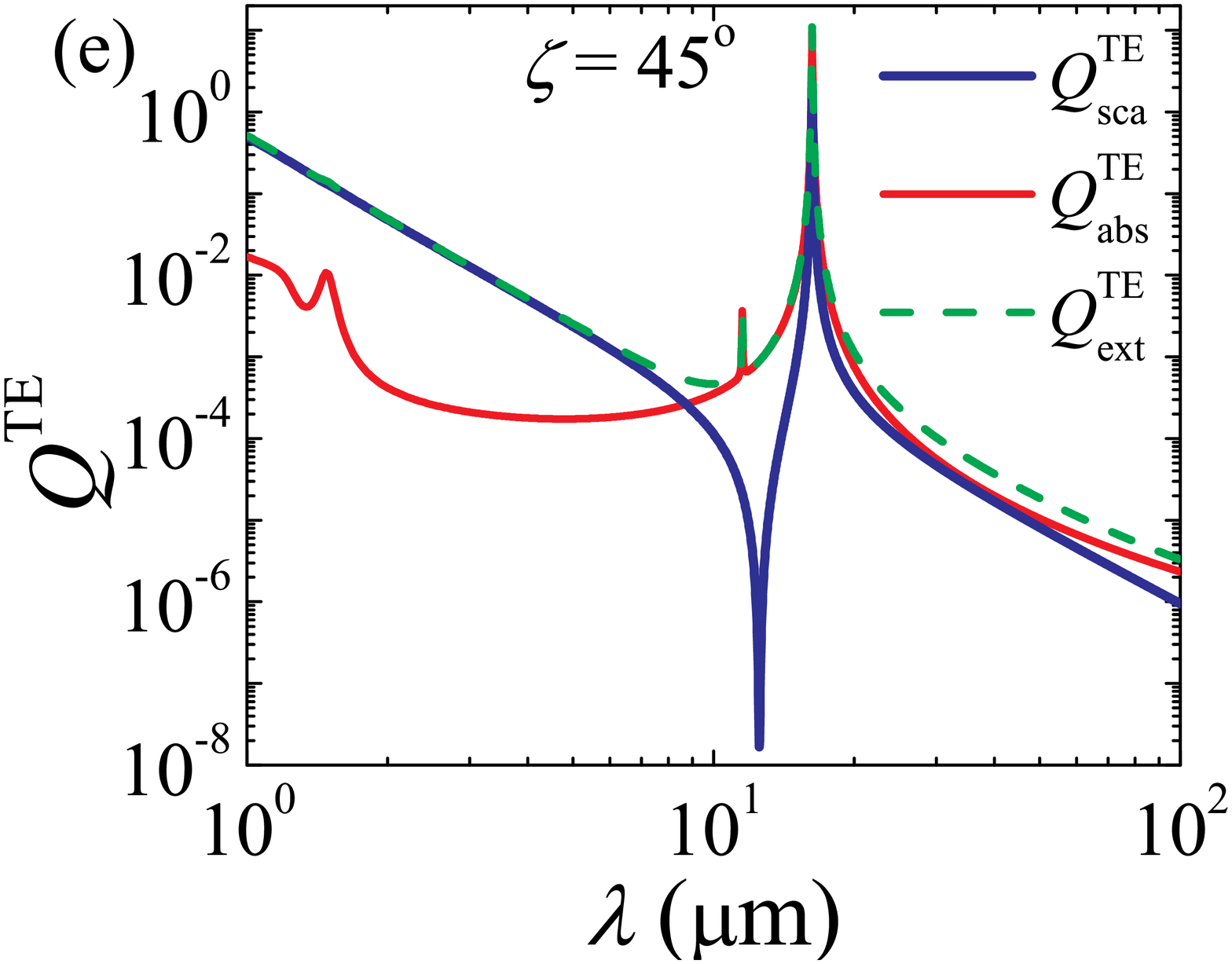}
\includegraphics[width=.66\columnwidth]{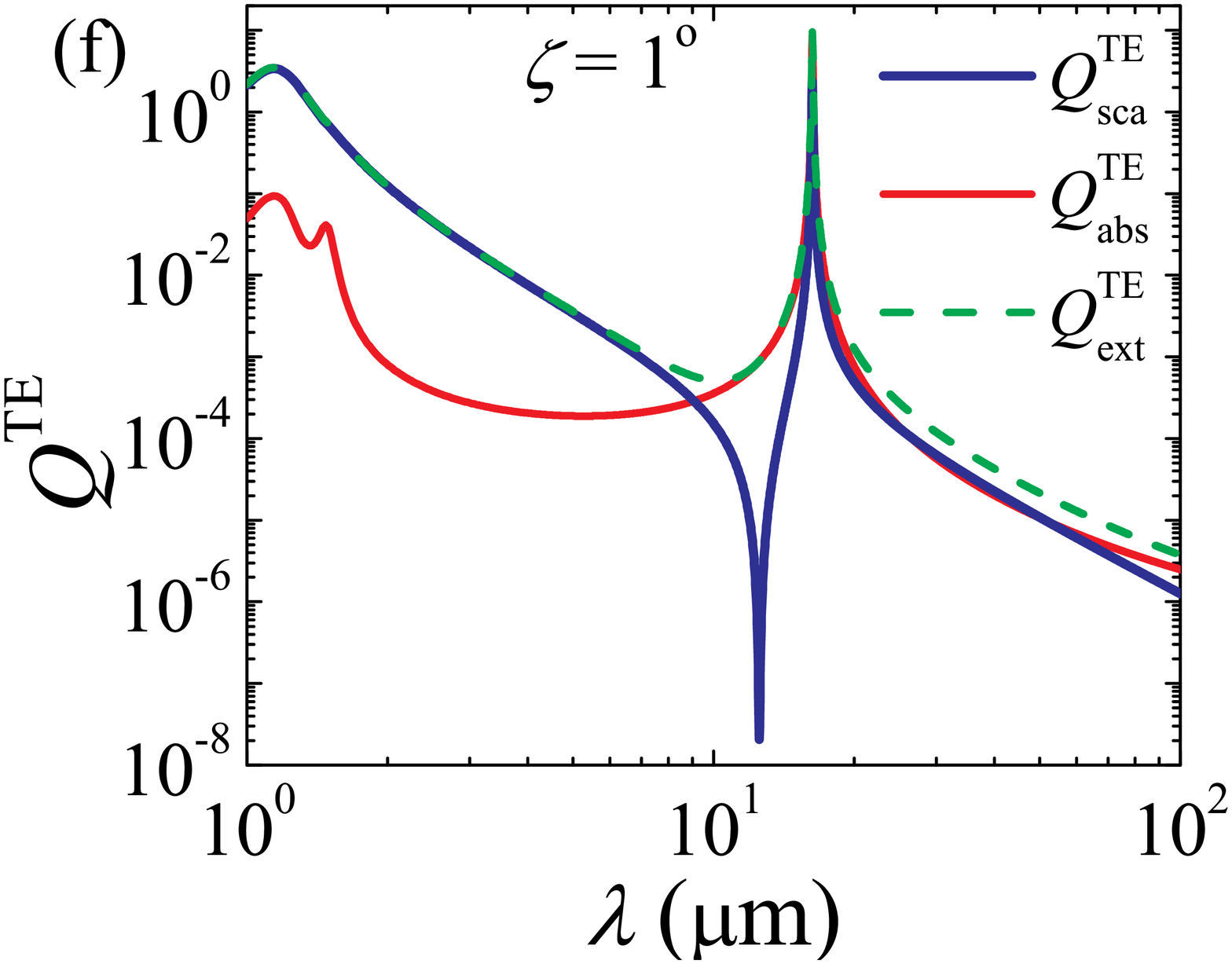}
\caption{
    Light scattering by a dielectric nanowire ($\varepsilon_{1}=3.9\varepsilon_0$) of radius $a=100$~nm coated with a graphene monolayer ($\mu_{\rm c}=0.5$~eV, $T=300$~K).
    The system is illuminated with TM ($\mathbf{H}\perp\hat{\mathbf{z}}$) or TE ($\mathbf{E}\perp\hat{\mathbf{z}}$) polarized plane waves, with $\zeta$ being the angle between the cylinder axis and the wavevector.
    The plots show the scattering ($Q_{\rm sca}$), absorption $(Q_{\rm abs})$, and extinction $(Q_{\rm ext})$ efficiencies for TM and TE-polarized waves as a function of the wavelength $\lambda$.
    For the TM mode, one has (a) $\zeta=90^{\rm o}$, (b) $\zeta=45^{\rm o}$, and (c) $\zeta=1^{\rm o}$.
    For the TE mode, one has (d) $\zeta=90^{\rm o}$, (e) $\zeta=45^{\rm o}$, and (f) $\zeta=1^{\rm o}$.
    A Fano-like resonance in $Q_{\rm sca}$ is obtained for both illumination schemes, with Fano dips at $\lambda_{\rm inv}^{\rm TM}\approx8.88~\mu$m (TM mode only) and $\lambda_{\rm inv}\approx 12.55~\mu$m, and localized surface plasmon resonance $(\ell=1)$ at $\lambda_{\rm res}\approx 16.25~\mu$m.
    The small peak that appears in $Q_{\rm abs}$ at $\lambda\approx11.52~\mu$m is associated with the quadrupole resonance ($\ell=2$).
}\label{fig2}
\end{figure*}

Within the full-wave Lorenz-Mie theory, the optical properties of the graphene-coated nanowire is described by the scattering, extinction, and absorption efficiencies given by Eqs.~(\ref{Qsca})--(\ref{Qabs}), respectively.
For a while, let us consider a graphene monolayer permittivity with fixed parameters $t_{\rm g}=0.5$~nm, $T=300$~K, $\mu_{\rm c}=0.5$~eV, and $\hbar\gamma=0.1$~meV~\cite{Naserpour_SciRep7_2017}.
 Here, we only consider the infrared regime where the surface plasmon is unimpeded by
surface optical phonons supported in graphene/dielectric structures~\cite{Basov_Nature487_2012}.

Figure~\ref{fig2} shows the optical efficiencies calculated for a graphene-coated nanowire in free space illuminated with plane waves as a function of the wavelength.
The range of size parameters is $0.0063<kb<0.63$, i.e., the cylinder radius is subwavelength.
For a normally illuminated cylinder ($\zeta=90^{\rm o}$), there is a localized surface plasmon resonance at $\lambda_{\rm res}\approx 16.25~\mu$m for the TE mode [Fig.~\ref{fig2}(a)], which is absent in the TM mode [Fig.~\ref{fig2}(d)].
This resonance is associated with a nonvanishing component of electric field orthogonal to the cylindrical surface, as one can verify by comparing Figs.~\ref{fig2}(a), \ref{fig2}(b), and \ref{fig2}(c) (TM mode) with Figs.~\ref{fig2}(d), \ref{fig2}(e), and \ref{fig2}(f) (TE mode), respectively.
Indeed, for grazing angles $(\zeta\approx 0)$, both TM and TE modes provide the same efficiencies since the incoming electric field is approximately orthogonal to the cylinder axis for the TM mode.
This is shown in Figs.~\ref{fig2}(c) and \ref{fig2}(f).

The localized surface plasmon resonance and the scattering antiresonances observed in Fig.~\ref{fig2} can be explained by the Lorenz-Mie coefficients~\cite{Arruda_JOpt14_2012,Arruda_JOpSocAmA31_2014,Arruda_PhysRevA94_2016,Arruda_PhysRevA96_2017}.
In particular, for subwavelength-diameter cylinders ($kb\ll1$), only the lower electromagnetic modes $\ell=0$ and $\ell=\pm1$ contribute to the extinction~\cite{Bohren_Book_1983}.
For the sake of simplicity, let us consider the normal incidence ($\zeta=90^{\rm o}$) and a nonmagnetic cylinder $(\mu_1=\mu_2=\mu_0)$, which leads to $a_0^{\rm TE}=a_1^{\rm TM}=b_1^{\rm TM}=b_1^{\rm TE}=0$.
Note that the decay channels $\ell=\pm1$ are degenerate since the cylinder is made of isotropic materials~\cite{Arruda_PhysRevA94_2016}.
In the Rayleigh limit, we have the scattering efficiencies $Q_{\rm sca}^{\rm TM}\approx 2|b_0^{\rm TM}|^2/kb$ and $Q_{\rm sca}^{\rm TE}\approx 4|a_1^{\rm TE}|^2/kb$, where the nonvanishing scattering coefficients are~\cite{Arruda_JOSA32_2015}
\begin{subequations}
\begin{align}
b_0^{\rm TM}\approx -\imath\frac{\pi}{4}(kb)^2\left(\frac{\varepsilon_{\rm eff}^{||} - \varepsilon_0}{\varepsilon_0}\right) + \mathcal{O}[(kb)^4],\label{b0-TM}\\
a_1^{\rm TE} \approx -\imath\frac{\pi}{4}(kb)^2\left(\frac{\varepsilon_{\rm eff}^{\perp} - \varepsilon_0}{\varepsilon_{\rm eff}^{\perp} + \varepsilon_0}\right)+ \mathcal{O}[(kb)^4].\label{a1-TE}
\end{align}
\end{subequations}
For a coated cylinder, the Maxwell-Garnett effective permittivities are
\begin{subequations}
\begin{align}
\varepsilon_{\rm eff}^{||}&=S^2\varepsilon_1 + (1-S^2)\varepsilon_2,\\
\varepsilon_{\rm eff}^{\perp}&=\frac{\varepsilon_2\left[(1+S^2)\varepsilon_1 + (1-S^2)\varepsilon_2\right]}{(1-S^2)\varepsilon_1+(1+S^2)\varepsilon_2},\label{eps-perp}
\end{align}
\end{subequations}
with $S\equiv a/b$ being the thickness ratio.
The effective permittivities $\varepsilon_{\rm eff}^{||}$ and $\varepsilon_{\rm eff}^{\perp}$ are related to electric polarizability of the cylinder for electric field parallel or orthogonal to the cylinder axis, respectively.
For a graphene-coated nanowire, one has $\varepsilon_1=\varepsilon_{\rm d}$, $\varepsilon_2=\varepsilon_{\rm g}(\omega)$, and $S=1- t_{\rm g}/b\approx1$, since the graphene monolayer is atomically thin.
This leads to the effective permittivities
\begin{subequations}
\begin{align}
\varepsilon_{\rm eff}^{||}(\omega)&\approx\varepsilon_{\rm d} + 2\frac{t_{\rm g}}{b}\varepsilon_{\rm g}(\omega),\label{eps-eff-para}\\ \varepsilon_{\rm eff}^{\perp}(\omega)&\approx \varepsilon_{\rm d} + \frac{t_{\rm g}}{b}\varepsilon_{\rm g}(\omega),\label{eps-eff-perp}
\end{align}
\end{subequations}
which agrees with Ref.~\cite{Naserpour_SciRep7_2017}.
According to Eq.~(\ref{a1-TE}), a strong localized surface plasmon resonance of TE waves occurs when ${\rm Re}[\varepsilon_{\rm eff}^{\perp}(\omega_{\rm res})]=-\varepsilon_0$ and ${\rm Im}[\varepsilon_{\rm eff}^{\perp}(\omega_{\rm res})]\ll\varepsilon_0$.
This can be verified not only for TE waves but also for oblique incidence of TM waves, see Figs.~\ref{fig2}(b)--\ref{fig2}(f).
In addition, from Eq.~(\ref{b0-TM}), a bulk plasmon excitation may also occur for TM waves when ${\rm Re}[\varepsilon_{\rm eff}^{||}(\omega_{\rm res})]=0$ and ${\rm Im}[\varepsilon_{\rm eff}^{\perp}(\omega_{\rm res})]\gg\varepsilon_0$, which is not the case for our set of parameters.
Conversely, the plasmonic cloaking of the dielectric cylinder by the graphene monolayer occurs when $Q_{\rm sca}\approx 0$, i.e., ${\rm Re}[\varepsilon_{\rm eff}^{||}(\omega_{\rm inv}^{\rm TM})]=\varepsilon_0$ for normal incidence of TM waves [Fig.~\ref{fig2}(a)] and ${\rm Re}[\varepsilon_{\rm eff}^{\perp}(\omega_{\rm inv}^{\rm TE})]=\varepsilon_0$ for TM waves at grazing angles [Fig.~\ref{fig2}(c)] or TE waves [Figs.~\ref{fig2}(d)--\ref{fig2}(f)].
Due to the presence of factor 2 in Eq.~(\ref{eps-eff-para}), which is absent in Eq.~(\ref{eps-eff-perp}), we have $\omega_{\rm inv}^{\rm TM}\not=\omega_{\rm inv}^{\rm TE}$.

To  calculate analytically the frequencies associated with localized surface plasmon resonances and the plasmonic cloaking of the dielectric cylinder studied in Fig.~\ref{fig2}, we need a simplified model of graphene conductivity.
For moderate frequencies and large doping, the intraband conductivity $\sigma_{\rm intra}$ dominates the contribution to graphene permittivity.
 Imposing $\mu_{\rm c}\gg k_{\rm B}T$ in Eq.~(\ref{sigma-intra}), it follows that $\sigma\approx\sigma_{\rm intra} \approx \imath e^2\mu_{\rm c}/\pi\hbar^2(\omega+\imath\gamma)$.
For our set of parameters, this Drude-like model of graphene conductivity is valid in the far-infrared frequencies and beyond $(10~\mu{\rm m}<\lambda\ll 12.4~{\rm mm})$, as discussed in Ref.~\cite{Naserpour_SciRep7_2017}.

Let us consider the condition for localized surface plasmon resonance and plasmonic cloaking: ${\rm Re}[\varepsilon_{\rm eff}^{\perp}(\omega_{\pm})]=\pm\varepsilon_0$, where $\omega_{\rm res}=\omega_+$ and $\omega_{\rm inv}=\omega_-$.
Using the Drude-like model of graphene permittivity for $\omega\gg\gamma$ and Eq.~(\ref{eps-eff-perp}), we obtain~\cite{Naserpour_SciRep7_2017}
\begin{align}
\omega_{\pm} = \sqrt{\frac{e^2\mu_{\rm c}}{\pi\hbar^2b(\varepsilon_1\pm\varepsilon_0)}},\label{w-plus}
\end{align}
where we have considered $\varepsilon_1=\varepsilon_{\rm d}>\varepsilon_0$.
Thus, in Fig.~\ref{fig2}(d), we have $\lambda_{\rm res}\approx 2\pi c/\omega_+$ and $\lambda_{\rm inv}\approx 2\pi c/\omega_-$.
Note that $Q_{\rm sca}$ as a function of frequency exhibits a Fano line shape.
Indeed, for $|\omega_+-\omega_-|\ll\omega$, one can verify that
\begin{align}
Q_{\rm sca}^{\rm TE}\propto\ \frac{\left(\omega^2-\omega_{-}^2\right)^2+(\omega\gamma)^2}{\left(\omega^2-\omega_{+}^2\right)^2+(\omega\gamma)^2}\approx\frac{(F\gamma/2+\omega-\omega_{+})^2}{(\omega-\omega_+)^2+(\gamma/2)^2},\label{Qsca-TE-app}
\end{align}
where $F=(\omega_+-\omega_-)/(\gamma/2)\propto \sqrt{\mu_{\rm c}}$ is the Fano asymmetry parameter.
This Fano resonance occurs due to the interference between narrow, localized surface plasmon resonances excited in the surface of the graphene monolayer and a broad Lorenz-Mie resonance of the dielectric nanowire.
In particular, one can verify that the cloaking frequency for normal incidence of TM waves is simply $\omega_{\rm inv}^{\rm TM}=\sqrt{2}\omega_-$, i.e., $\lambda_{\rm inv}=\sqrt{2}\lambda_{\rm inv}^{\rm TM}$.
In fact, from Figs.~\ref{fig2}(a) and \ref{fig2}(c), one has $\lambda_{\rm inv}/\lambda_{\rm inv}^{\rm TM}\approx \sqrt{2}$.

The plasmonic Fano resonance is strongly dependent on the local dielectric environment and the geometry of the system, even for subwavelength structures~\cite{Kivshar_RevModPhys82_2010}.
This means that the overall scattering response depends on the cross sectional shape of the nanowire.
In fact, the breaking of the rotational symmetry is expected to affect the Fano line shape for thick layers, leading, e.g., to multiple Fano resonances or the elimination of the Fano dip in light scattering~\cite{Zayats_OptExp21_2013}.
All the discussion above assumes a nanowire with cylindrical geometry, i.e., rotational symmetry.
A precise description on how arbitrary cross sectional shapes of nano- or microwires would affect the Fano resonance using graphene coatings is a subject for another study and will be investigated elsewhere.
In particular, the spectral position of the enhancement and/or suppression of light scattering can be tuned by varying the volume ratio of plasmonic coating and dielectric core, see Eqs.~(\ref{eps-eff-para}) and (\ref{eps-eff-perp}).
Since the thickness $t_{\rm g}$ of the graphene monolayer is fixed, one could vary the position of the Fano resonance in the spectra by varying the diameter $a$ of the dielectric core and/or by considering multiple layers of graphene coatings.

\subsection{Spontaneous-emission rate near a graphene-coated nanowire}
\label{Fano}

 \begin{figure}[htbp!]
\includegraphics[width=\columnwidth]{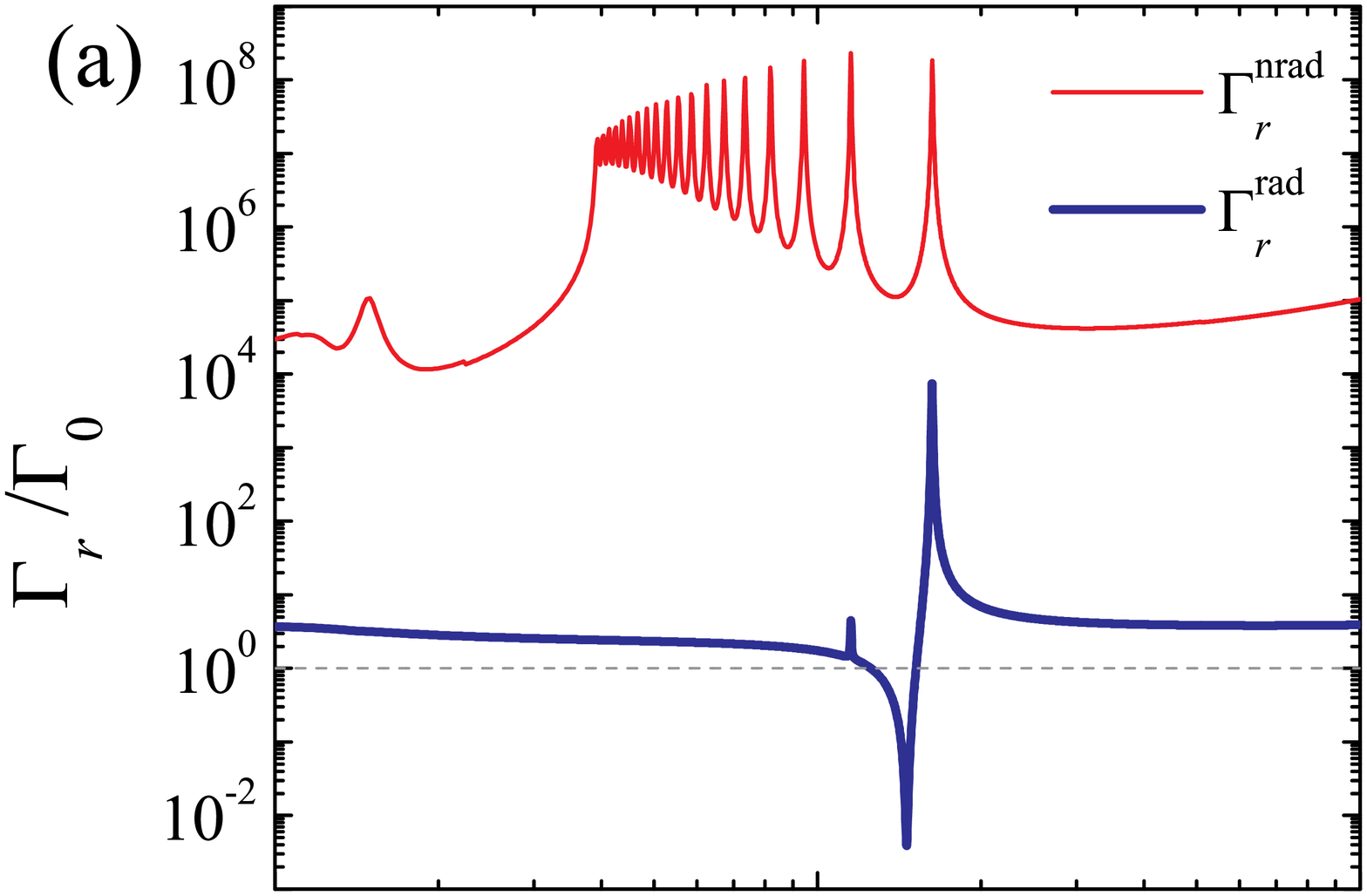}\vspace{-1.8cm}
\includegraphics[width=\columnwidth]{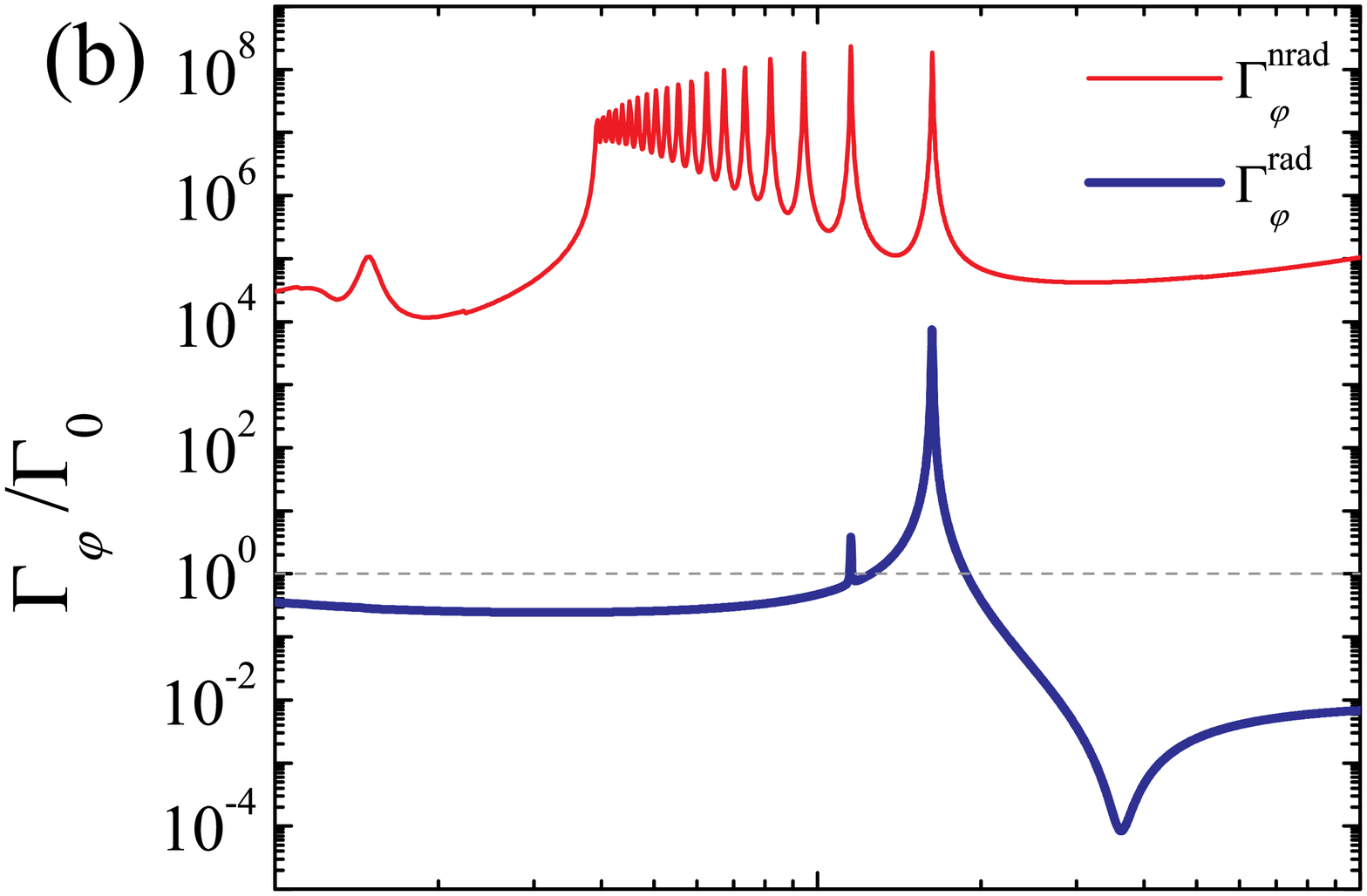}\vspace{-1.8cm}
\includegraphics[width=\columnwidth]{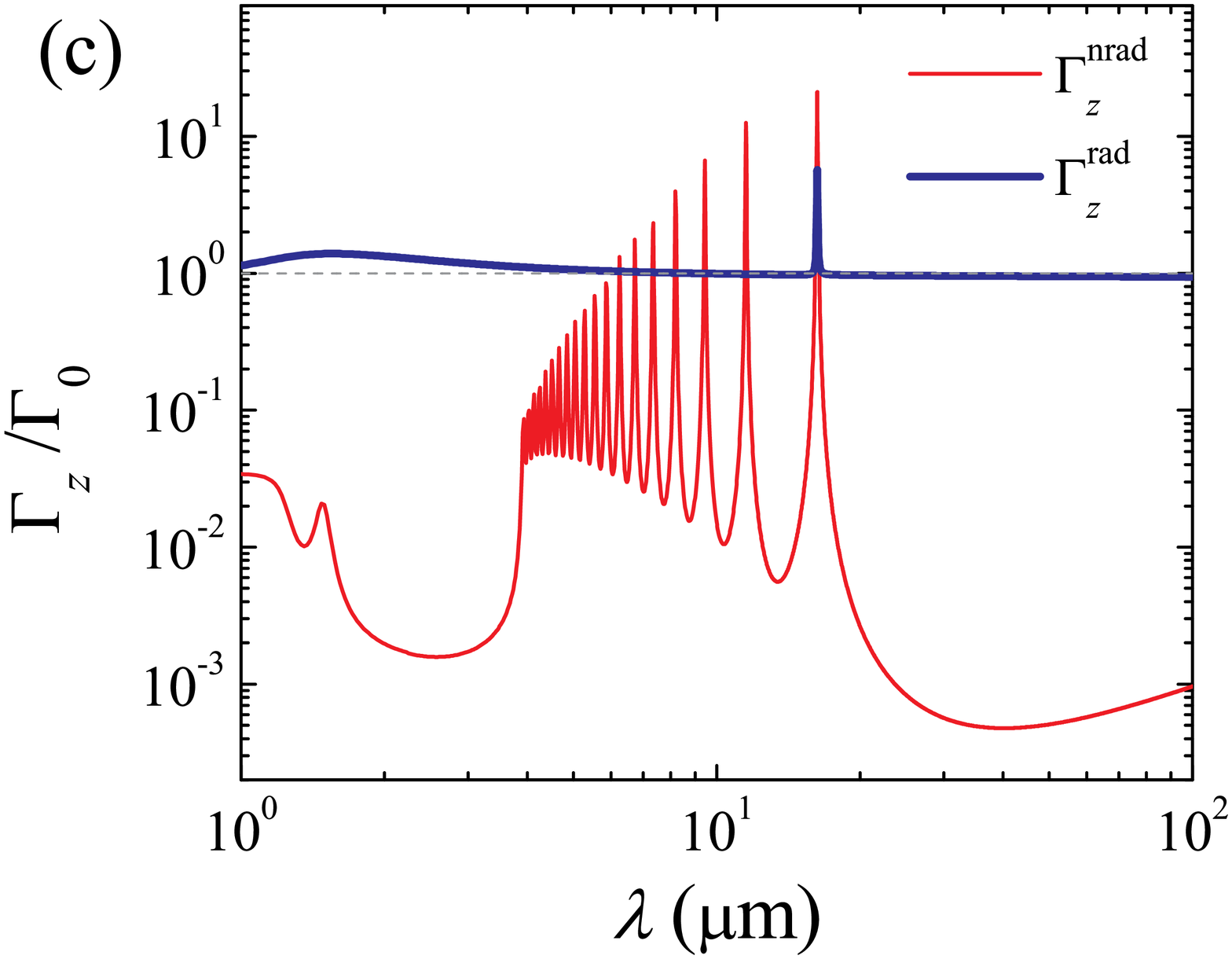}
\caption{Radiative ($\Gamma_{\mathbf{d}_0}^{\rm rad}$) and  nonradiative ($\Gamma_{\mathbf{d}_0}^{\rm nrad}$) decay rates (normalized to free space) associated with a point dipole emitter in the vicinity of a graphene-coated dielectric nanowire $(\mu_{\rm c}=0.5$~eV, $T=300$~K) of radius $a=100$~nm.
The distance between the dipole emitter and the graphene monolayer is $\Delta r=r'-b=5$~nm.
The plot shows the decay rates for dipole moments (a) $\mathbf{d}_0=d_0\hat{\mathbf{r}}$, (b) $\mathbf{d}_0=d_0\hat{\boldsymbol{\varphi}}$, and
(c) $\mathbf{d}_0=d_0\hat{\mathbf{z}}$.
The enhancement and suppression of $\Gamma_{\mathbf{d}_0}^{\rm rad}$ (thick line) are due to the excitation of localized surface plasmons.
Conversely, the peaks in $\Gamma_{\mathbf{d}_0}^{\rm nrad}$ (thin line) are mainly due to ohmic losses in association with guided modes within the nanobody.
}\label{fig3}
\end{figure}

 \begin{figure}[htbp!]
\includegraphics[width=\columnwidth]{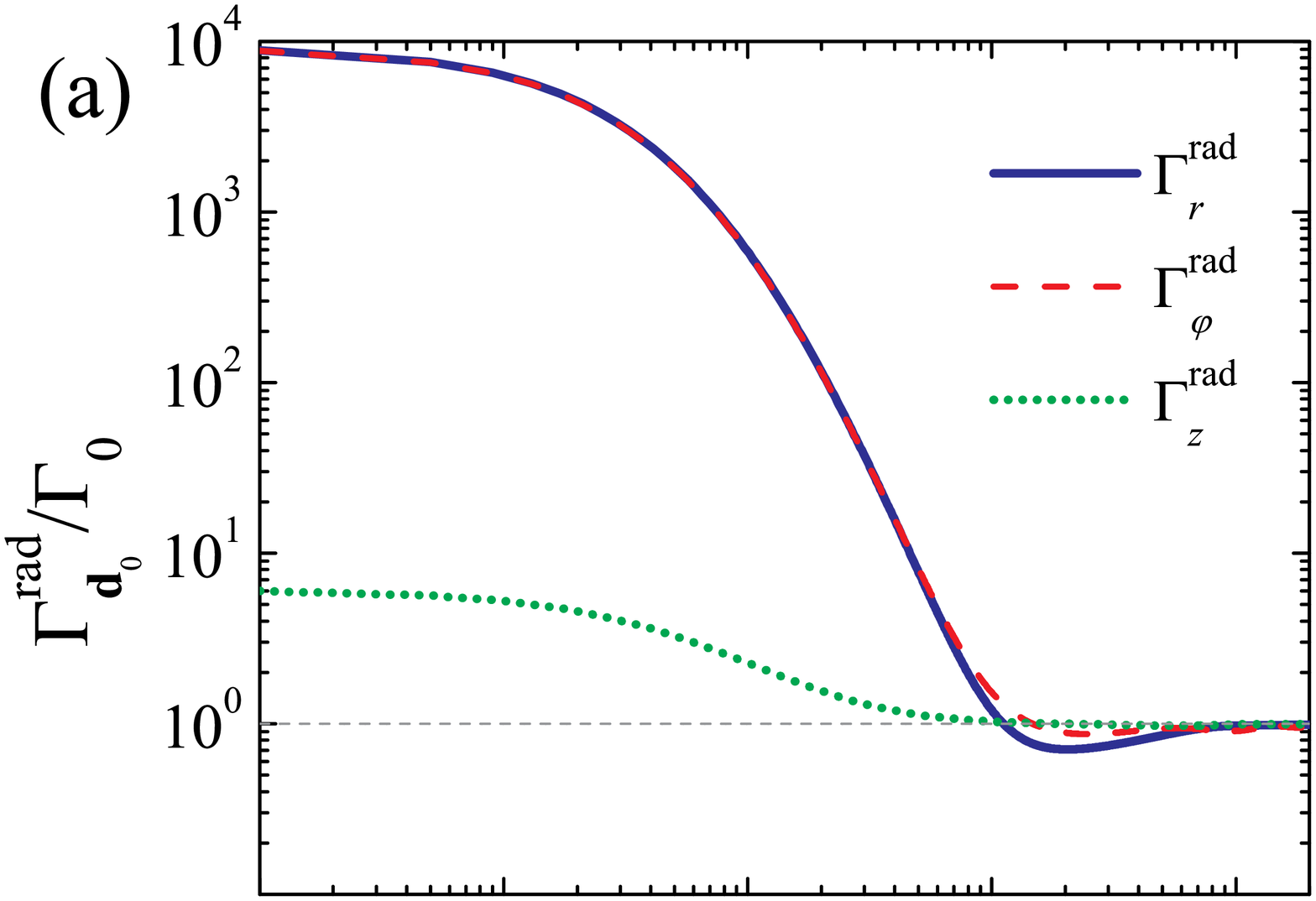}\vspace{-1.8cm}
\includegraphics[width=\columnwidth]{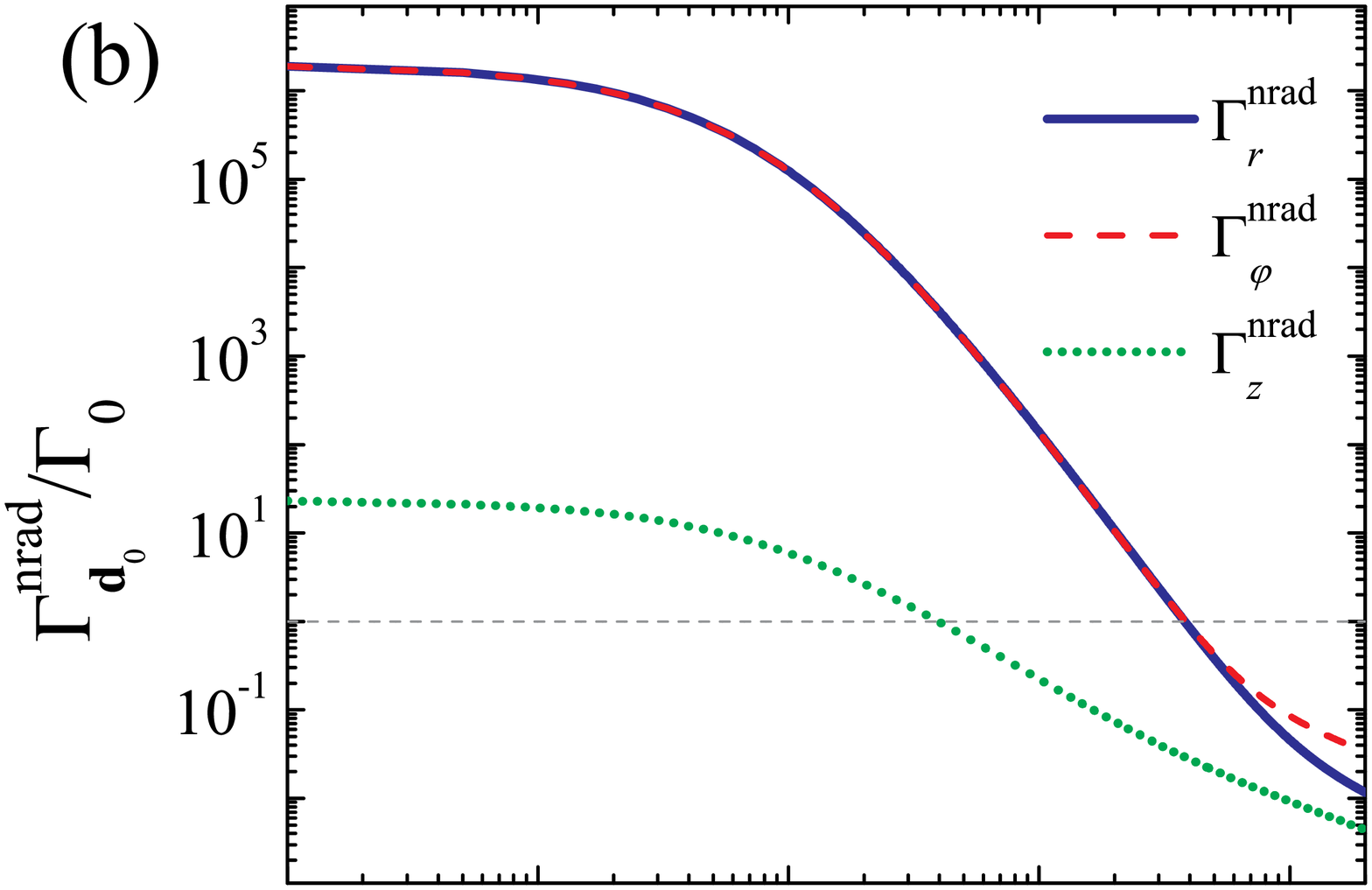}\vspace{-1.8cm}
\includegraphics[width=\columnwidth]{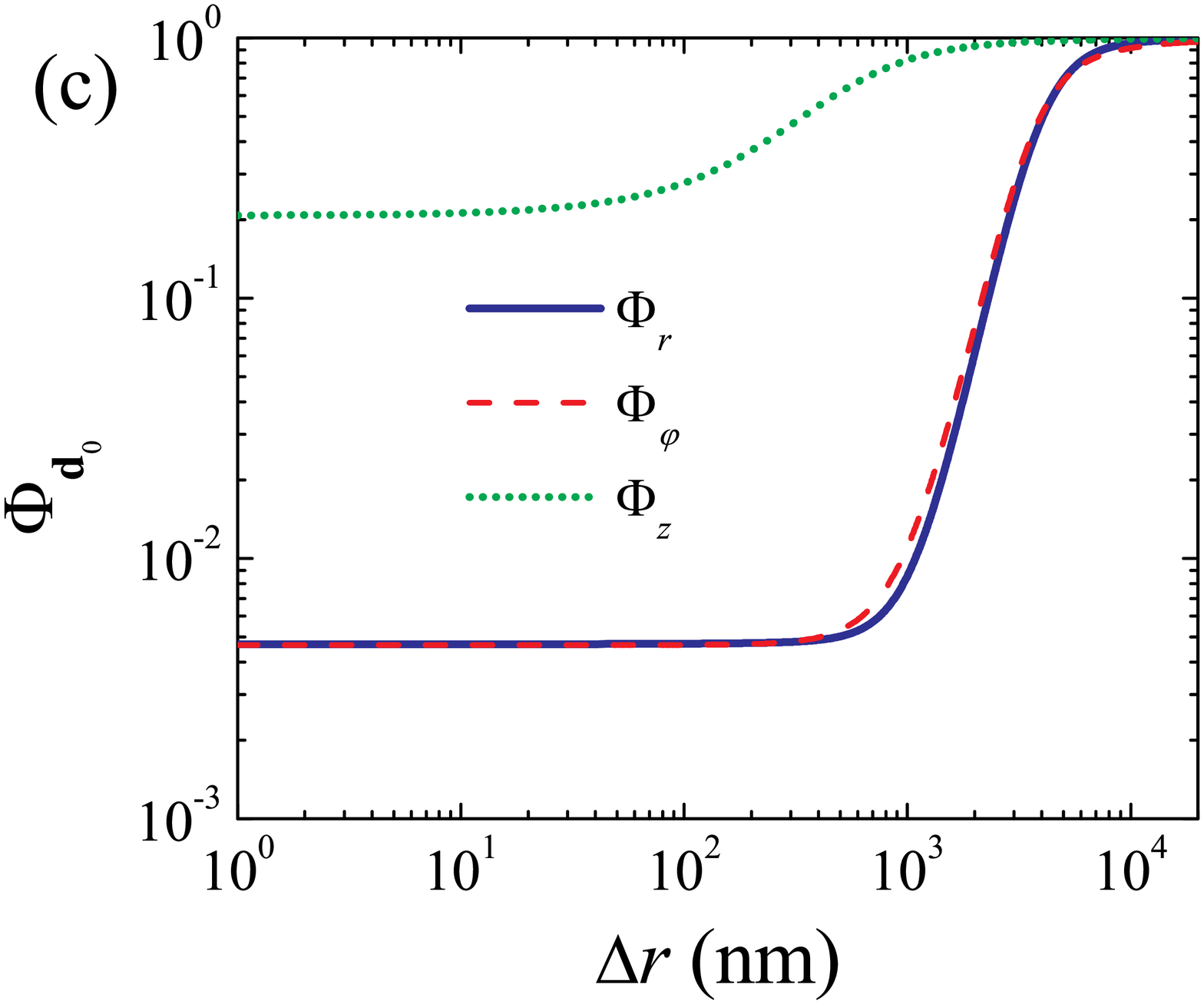}
\caption{Decay rates of a dipole emitter near a graphene-coated dielectric nanowire ($\mu_{\rm c}=0.5$~eV, $T=300$~K) of radius $a=100$~nm and permittivity $\varepsilon_{\rm d}=3.9\varepsilon_0$ as a function of the distance $\Delta r$ between emitter and cylinder.
The emission wavelength is $\lambda_{\rm res}\approx16.25~\mu$m, which corresponds to the localized plasmon resonance of the effective cylinder: ${\rm Re}[\varepsilon_{\rm eff}^{\perp}(\lambda_{\rm res})]\approx-\varepsilon_0$.
The plots show (a) radiative and (b) nonradiative decay rates normalized to free space for three dipole moment orientations, and (c) the corresponding radiation efficiency $\Phi_{\mathbf{d}_0}=\Gamma_{\mathbf{d}_0}^{\rm rad}/(\Gamma_{\mathbf{d}_0}^{\rm rad}+\Gamma_{\mathbf{d}_0}^{\rm nrad})$.
}\label{fig4}
\end{figure}

Based on the previous discussion, let us consider a point dipole emitter in the vicinity of a dielectric nanowire of radius $a=100$~nm coated with a graphene monolayer, which enters into Kubo's formula with parameters $\mu_{\rm c}=0.5$~eV, $T=300$~K, and $\hbar\gamma=0.1$~meV.
The dipole emitter is located at distance $\mathbf{r}'$ in a cylindrical coordinate system (Fig.~\ref{fig1}), and its corresponding radiative and nonradiative decay rates are given in Appendix~\ref{decay-rates}, see Eqs.~(\ref{Gamma-cylinder-r-rad})--(\ref{Gamma-cylinder-z-rad}) and Eqs.~(\ref{Gamma-cylinder-r-nrad})--(\ref{Gamma-cylinder-z-nrad}), respectively.
In the midinfrared, the pointlike emitter could be a nanoemitter such as an artificial atom or a quantum dot~\cite{Sweeny_NatPhotonics8_2014,Jiang_OptExp22_2014}, where higher-order electric interactions within the nanostructure can be neglected in comparison to electric dipole interactions.
In particular, it is desirable to have a finite distance between emitter and the plasmonic coating to reduce nonradiative contributions.
Typically, one considers a transparent dielectric spacer with small refractive index (e.g., a polymeric material) between emitter and the plasmonic structure, where the former is placed on top of the dielectric surface~\cite{Klimov_PhysRevA69_2004}.
This is a common procedure in biomedical applications, where the dielectric spacer can even enhance the radiation efficiency~\cite{Lin_SciRep5_2015}.
For the sake of simplicity, we consider both dipole emitter and graphene-coated nanowire in free space.

Figure~\ref{fig3} shows radiative and nonradiative contributions to the Purcell factor as a function of the emission wavelength.
The dipole emitter is located at distance $\Delta r=r'-b=5$~nm from the graphene surface.
In Figs.~\ref{fig3}(a)--\ref{fig3}(c), we consider three basic orientations of electric dipole moment according to Appendix~\ref{decay-rates}, respectively: $\mathbf{d}_0=d_0\hat{\mathbf{r}}$, $\mathbf{d}_0=d_0\hat{\boldsymbol{\varphi}}$, and
$\mathbf{d}_0=d_0\hat{\mathbf{z}}$.
Note that a similar signature as the one obtained in the scattering efficiency $Q_{\rm sca}$, Fig.~\ref{fig2}(d), appears in $\Gamma_r^{\rm rad}$ and $\Gamma_{\varphi}^{\rm rad}$ as a function of $\lambda$.
The maximum enhancement of the radiative decay rate $\Gamma_{\mathbf{d}_0}^{\rm rad}$, for all three basic orientations ($\Gamma_r^{\rm rad}\approx \Gamma_{\varphi}^{\rm rad}\approx10^4\Gamma_0$ and $\Gamma_z^{\rm rad}\approx 10\Gamma_0$), is achieved at the plasmon resonance frequency obtained from Eq.~(\ref{w-plus}), i.e., $\lambda_{\rm res}\approx16.25~\mu$m.
However, the suppression of the radiative decay rate that we see in Fig.~\ref{fig3}(a) ($\Gamma_r^{\rm rad}\approx 10^{-3}\Gamma_0$) and Fig.~\ref{fig3}(b) ($\Gamma_{\varphi}^{\rm rad}\approx 10^{-4}\Gamma_0$), which is related to the plasmonic cloaking of the dielectric nanowire $(Q_{\rm sca}\approx0)$, depends on the distance $\Delta r$ between emitter and cylinder, and hence cannot be predicted by the scattering efficiency alone.
In addition, the nonradiative decay rate $\Gamma_{\mathbf{d}_0}^{\rm nrad}$ shows a series of narrow peaks for ${\rm Re}[\varepsilon_{\rm eff}^{\perp}(\omega)]>0$, which are due to ohmic losses in the graphene monolayer and can also be associated with guided modes within the cylindrical nanobody.

 A first conclusion derived from Fig.~\ref{fig3} is that the total decay rate for a dipolar emitter with dipole moment oriented along the cylinder axis $(\Gamma_z)$ is much smaller than the decay rate of a dipole moment oriented orthogonal to the cylinder axis $(\Gamma_{r},\Gamma_{\varphi})$.
 For planar metallic surfaces, this effect is usually explained by the interaction between a real dipole and its dipole image.
 For an electric dipole moment orthogonal to a plasmonic surface, one obtains the enhancement of the optical response due to constructive interference with its corresponding dipole image.
 Conversely, for an electric dipole moment parallel to the plasmonic surface, the real dipole and its image are out-of-phase and the interference is destructive, thus suppressing the optical response.
 Even at the localized plasmon resonance wavelength $\lambda_{\rm res}\approx 16.25~\mu$m, the difference between the decay rates for dipoles with either orthogonal or parallel orientation is large, as can be seen in Fig.~\ref{fig4}.
 Finally, for the present set of parameters, there is no great variation on the radiation efficiency $\Phi_{\mathbf{d}_0}\equiv\Gamma_{\mathbf{d}_0}^{\rm rad}/(\Gamma_{\mathbf{d}_0}^{\rm rad}+\Gamma_{\mathbf{d}_0}^{\rm nrad})$ for $0<\Delta r<100$~nm.
 As expected, for $kr'\gg1$, one has $\Gamma_{\mathbf{d}_0}^{\rm rad}\to\Gamma_0$ and $\Gamma_{\mathbf{d}_0}^{\rm nrad}\to 0$, i.e., $\Phi_{\mathbf{d}_0}\to1$.

\begin{figure}[htbp!]
\includegraphics[width=\columnwidth]{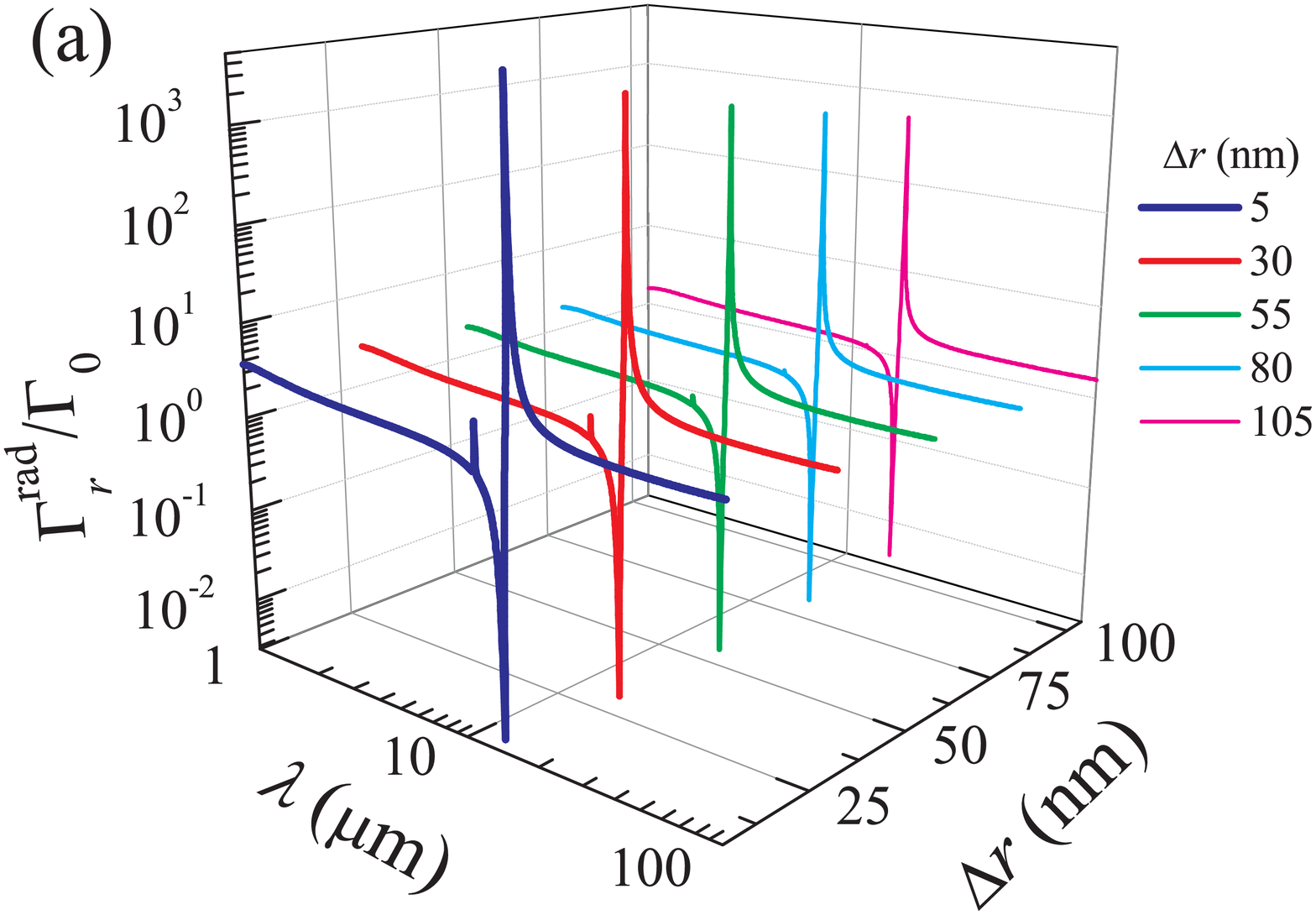}\vspace{-.5cm}
\includegraphics[width=\columnwidth]{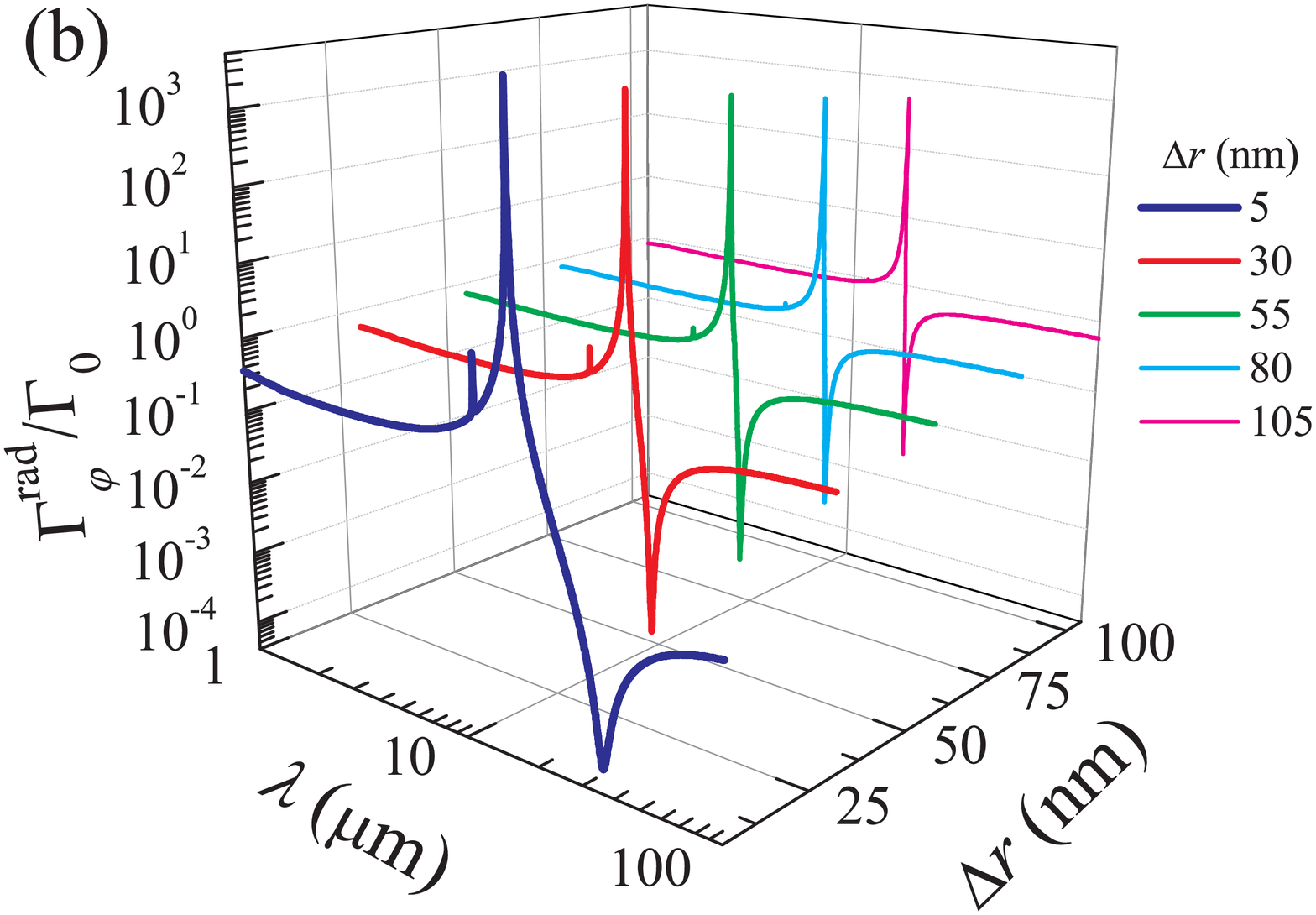}
\caption{
Radiative decay rates related to a dipole emitter in the vicinity of a graphene-coated dielectric nanowire ($\mu_{\rm c}=0.5$~eV, $T=300$~K).
The dielectric core has radius $a=100$~nm and permittivity $\varepsilon_{\rm d}=3.9\varepsilon_0$.
The plots show (a) $\Gamma_r^{\rm rad}$ and (b) $\Gamma_{\varphi}^{\rm rad}$ as a function of the emission wavelength and the distance $\Delta r$ between emitter and cylinder.
The enhancement due to the localized surface plasmon resonance occurs at $\lambda_{\rm res}\approx 16.25~\mu$m for any $\Delta r$.
The suppression of radiative emission, however, is dependent of $\Delta r$ due to guided mode contributions near the cylindrical surface.
}\label{fig5}
\end{figure}

 In Figs.~\ref{fig3}(a) and \ref{fig3}(b), the radiative decay rate exhibits a Fano-like resonance in the near field due the interference between a localized surface plasmon resonance and a broad dipole resonance acting as a background.
 The position of the maximum enhancement of $\Gamma_{\mathbf{d}_0}^{\rm rad}$ coincides with the maximum of the scattering efficiency $Q_{\rm sca}^{\rm TE}$ at ${\rm Re}[\varepsilon_{\rm eff}^{\perp}(\omega_+)]\approx-\varepsilon_0$.
 This is explained by the relation between the LDOS and the scattered electromagnetic fields, highlighted in Eq.~(\ref{definition}).
 To discuss the Fano dip in $\Gamma_r^{\rm rad}$ and $\Gamma_{\varphi}^{\rm rad}$, however, we need approximate expressions for the radiative decay rates.
 As can be seen in Fig.~\ref{fig5}, the Fano dip depends on the distance $\Delta r$ between emitter and cylinder, and this dependence is stronger for $\Gamma_{\varphi}^{\rm rad}$.
 In addition, the electric quadrupole contribution $(\ell=2)$, which appears as a small peak at $\lambda\approx11.52~\mu$m in Figs.~\ref{fig5}(a) and \ref{fig5}(b),  vanishes for $r'$ sufficiently larger than $b$.
 This higher-order mode contribution associated with a subwavelength-diameter nanobody is mainly related to near-field interactions between emitter and nanobody.

 Let us consider the limiting case $kr'\ll1$, i.e., the system composed by cylinder and emitter together is diameter-subwavelength, and hence can be described as a dipole-type system~\cite{Klimov_PhysRevA69_2004}.
 Assuming that the main contribution to the Purcell factor is achieved at $\zeta=90^{\rm o}$, and using the approximations $J_1(\rho)\approx \rho/2$ and $H_1^{(1)}(\rho)\approx-2\imath/\pi\rho$, one has from Eqs.~(\ref{Gamma-cylinder-r-rad})--(\ref{Gamma-cylinder-z-rad}), respectively,
 \begin{subequations}
\begin{align}
\frac{\Gamma_{r}^{\rm rad}(r',\omega)}{\Gamma_0} &\approx\left|1 + \left[\frac{\varepsilon_{\rm eff}^{\perp}(\omega) - \varepsilon_0}{\varepsilon_{\rm eff}^{\perp}(\omega)+\varepsilon_0}\right]\frac{b^2}{r'^2}\right|^2,\label{Gamma-r-app}\\
\frac{\Gamma_{\varphi}^{\rm rad}(r',\omega)}{\Gamma_0} &\approx\left|1 - \left[\frac{\varepsilon_{\rm eff}^{\perp}(\omega) - \varepsilon_0}{\varepsilon_{\rm eff}^{\perp}(\omega)+\varepsilon_0}\right]\frac{b^2}{r'^2}\right|^2,\label{Gamma-phi-app}\\
\frac{\Gamma_{z}^{\rm rad}(r',\omega)}{\Gamma_0} &\approx1,\label{Gamma-z-app}
\end{align}
\end{subequations}
where $\varepsilon_{\rm eff}^{\perp}(\omega)$ is calculated by Eq.~(\ref{eps-perp}).

The electrostatics analysis that leads to Eqs.~(\ref{Gamma-r-app})--(\ref{Gamma-z-app}) is discussed in detail in Ref.~\cite{Klimov_PhysRevA69_2004}, assuming a homogeneous nanofiber with ${\rm Re}[\varepsilon_{\rm eff}^{\perp}(\omega)]=\varepsilon>0$.
In fact, note that Eq.~(\ref{Gamma-z-app}) is not valid at the localized surface plasmon resonance, see Fig.~\ref{fig3}(c).
We can show, however, that this approximation can also be applied for ${\rm Re}[\varepsilon_{\rm eff}^{\perp}(\omega)]<0$ (plasmonic cylinder) as long as $kb\ll kr'$.
The main reason is that Eqs.~(\ref{Gamma-r-app})--(\ref{Gamma-z-app}) are not taking into account the guided mode contribution (bulk plasmons) related to $\ell=0$ and higher orders $(\ell\geq2)$.
Since the decay rates of guided modes are exponentially small as a function of $k\Delta r$~\cite{Girard_JOpt18_2016}, their influence on the radiative contribution can be neglected in the far field.

\begin{figure}[htbp!]
\includegraphics[width=\columnwidth]{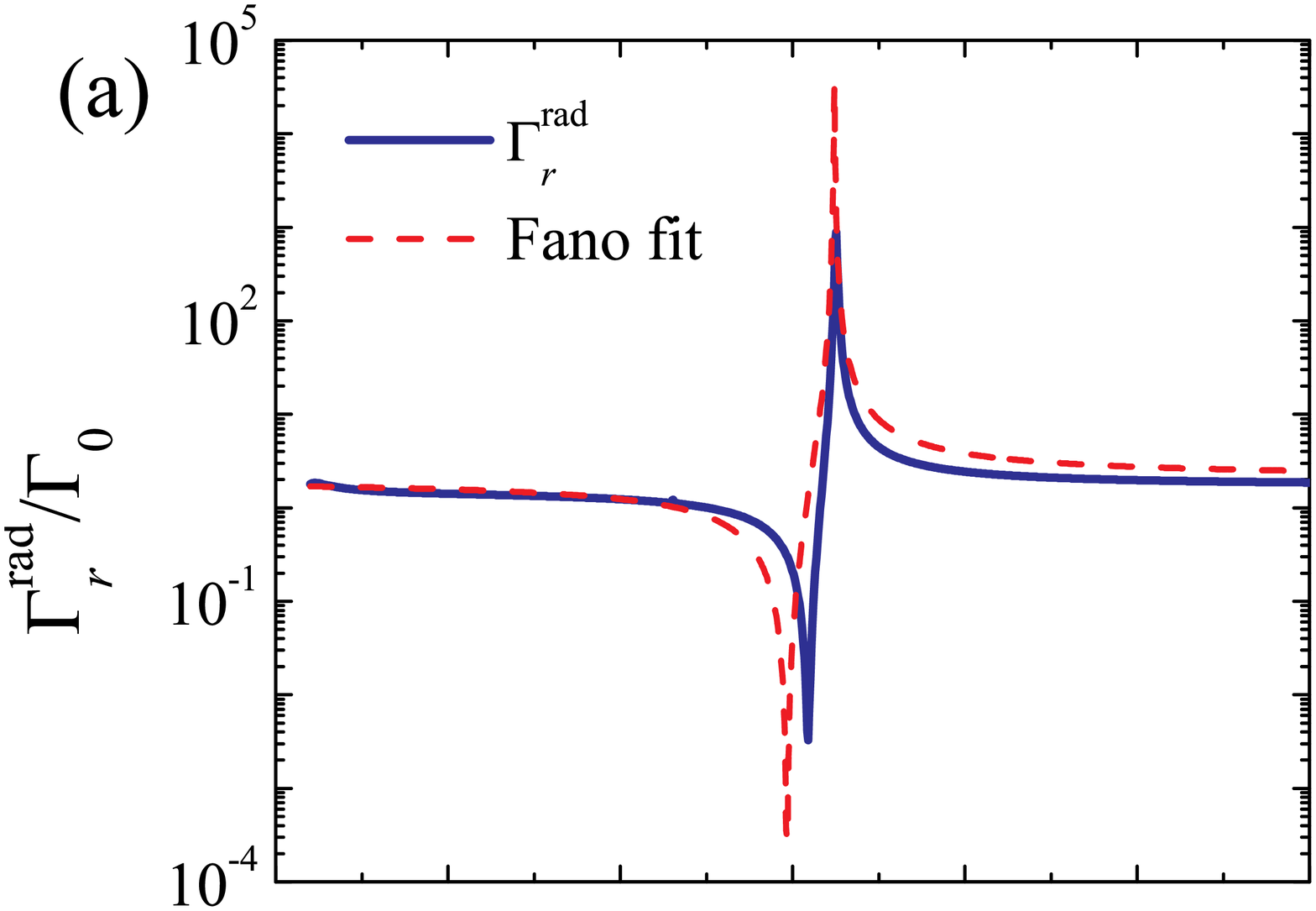}\vspace{-1.8cm}
\includegraphics[width=\columnwidth]{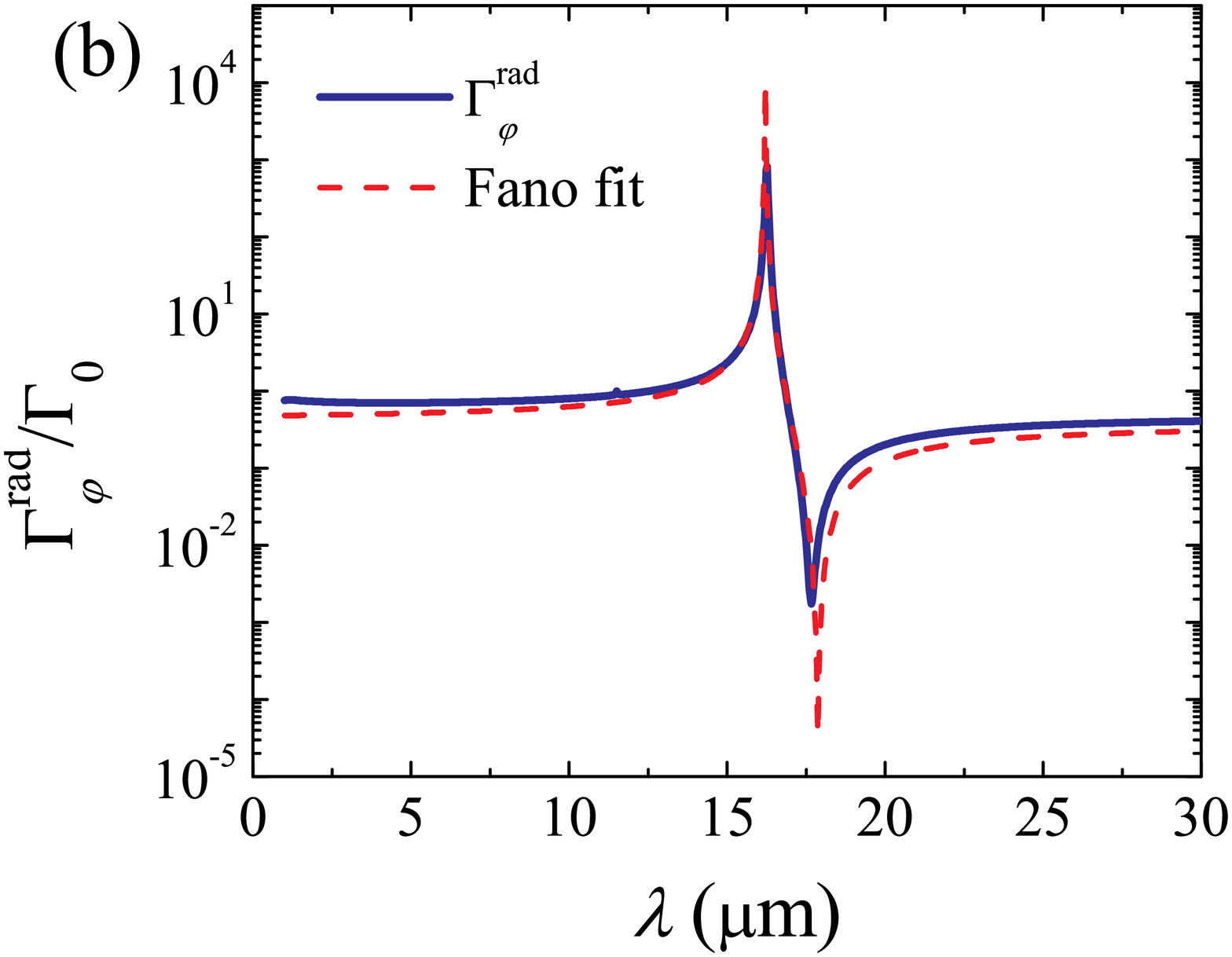}
\caption{
Radiative decay rates for a point dipole emitter near a graphene-coated dielectric nanowire ($\mu_{\rm c}=0.5$~eV, $T=300$~K).
The dielectric core has radius $a=100$~nm and permittivity $\varepsilon_{\rm d}=3.9\varepsilon_0$.
The plots show $\Gamma_r^{\rm rad}$ and $\Gamma_{\varphi}^{\rm rad}$ for a distance $\Delta r = r'-b=80$~nm between emitter and cylinder as a function of $\lambda$.
The dotted lines (Fano fit) are approximate curves calculated from Eqs.~(\ref{Gamma-rad-r-app}) and (\ref{Gamma-rad-phi-app}).
}\label{fig6}
\end{figure}

\begin{figure}[htbp!]
\includegraphics[width=\columnwidth]{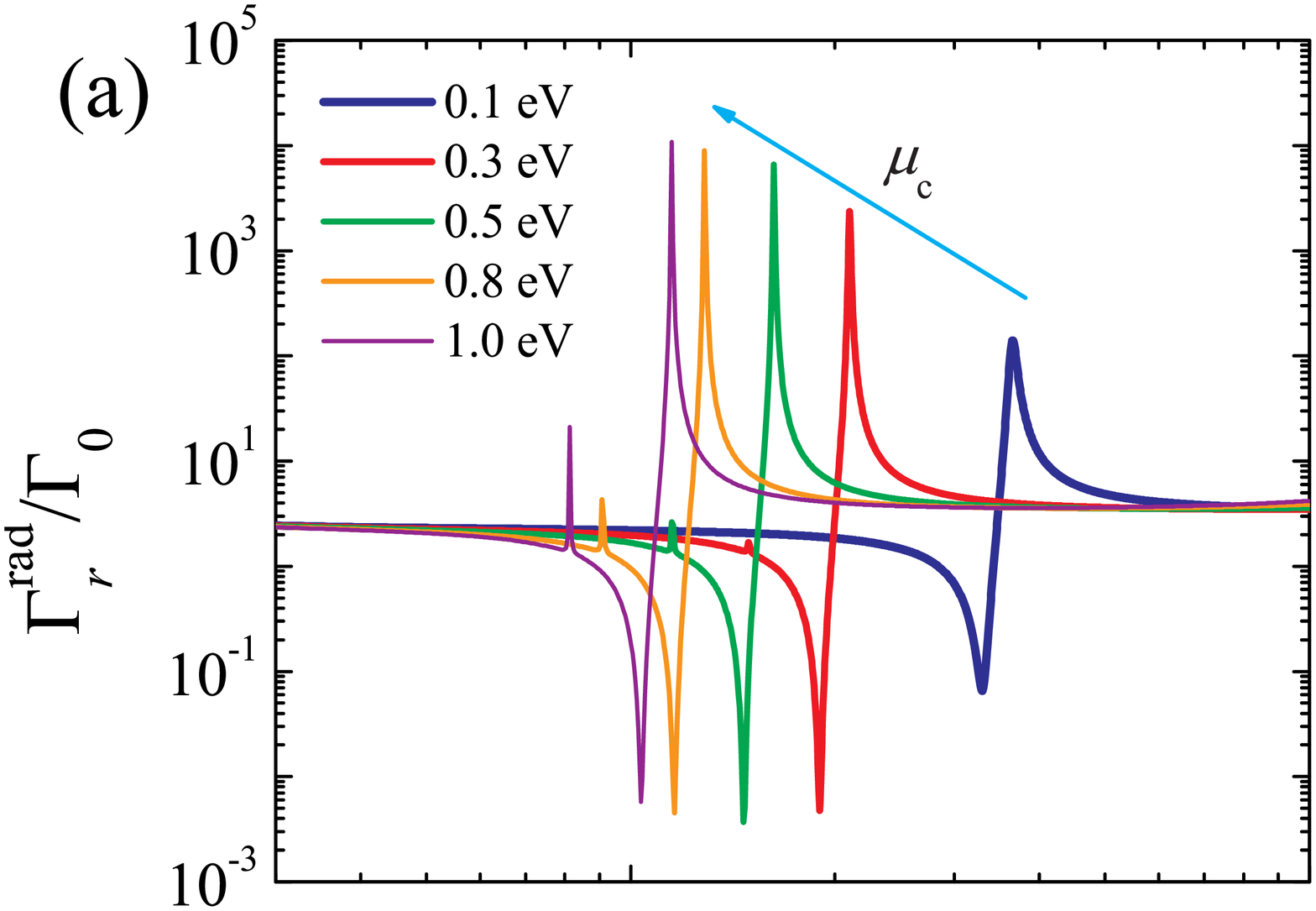}\vspace{-1.8cm}
\includegraphics[width=\columnwidth]{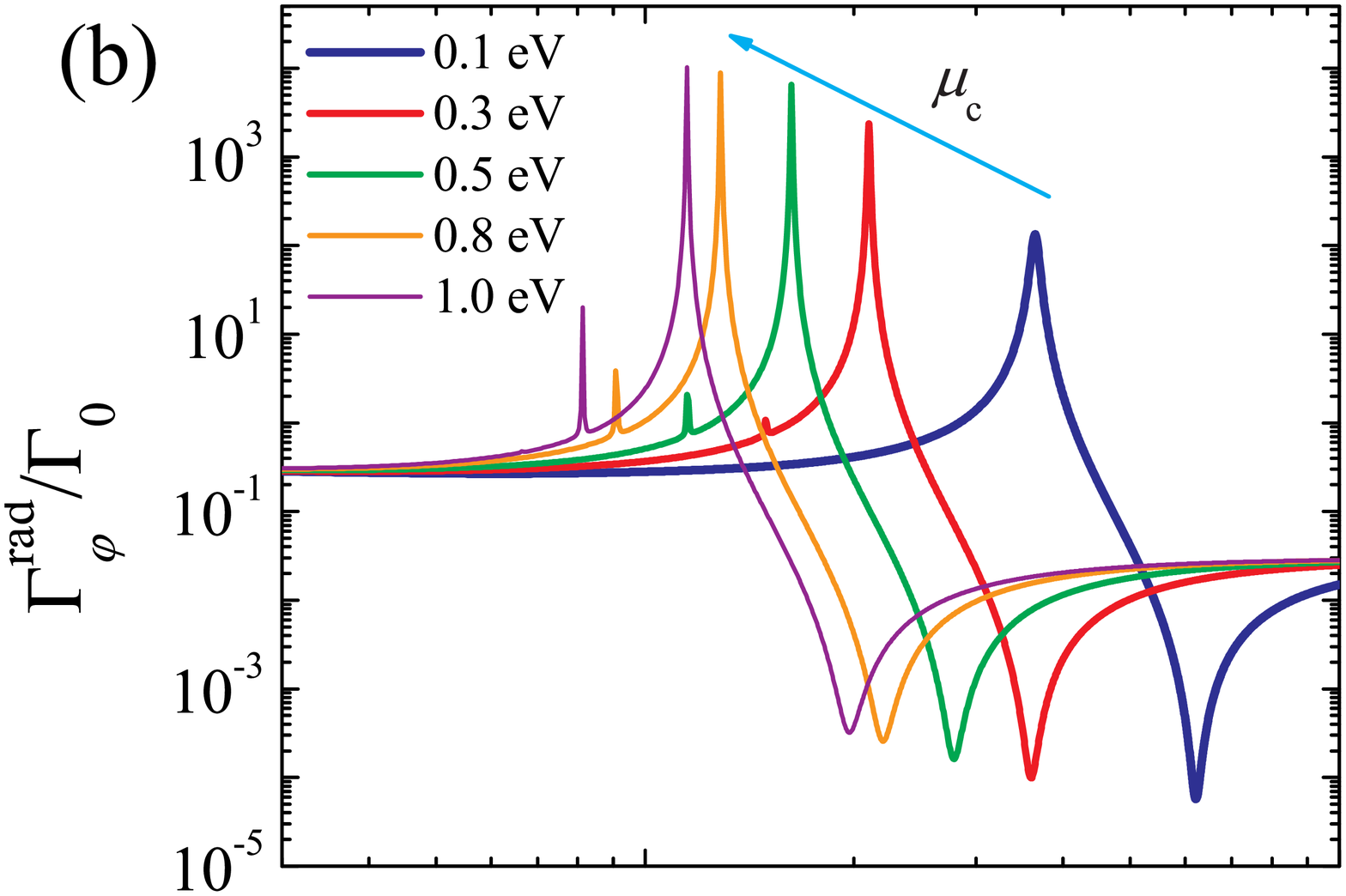}\vspace{-1.8cm}
\includegraphics[width=\columnwidth]{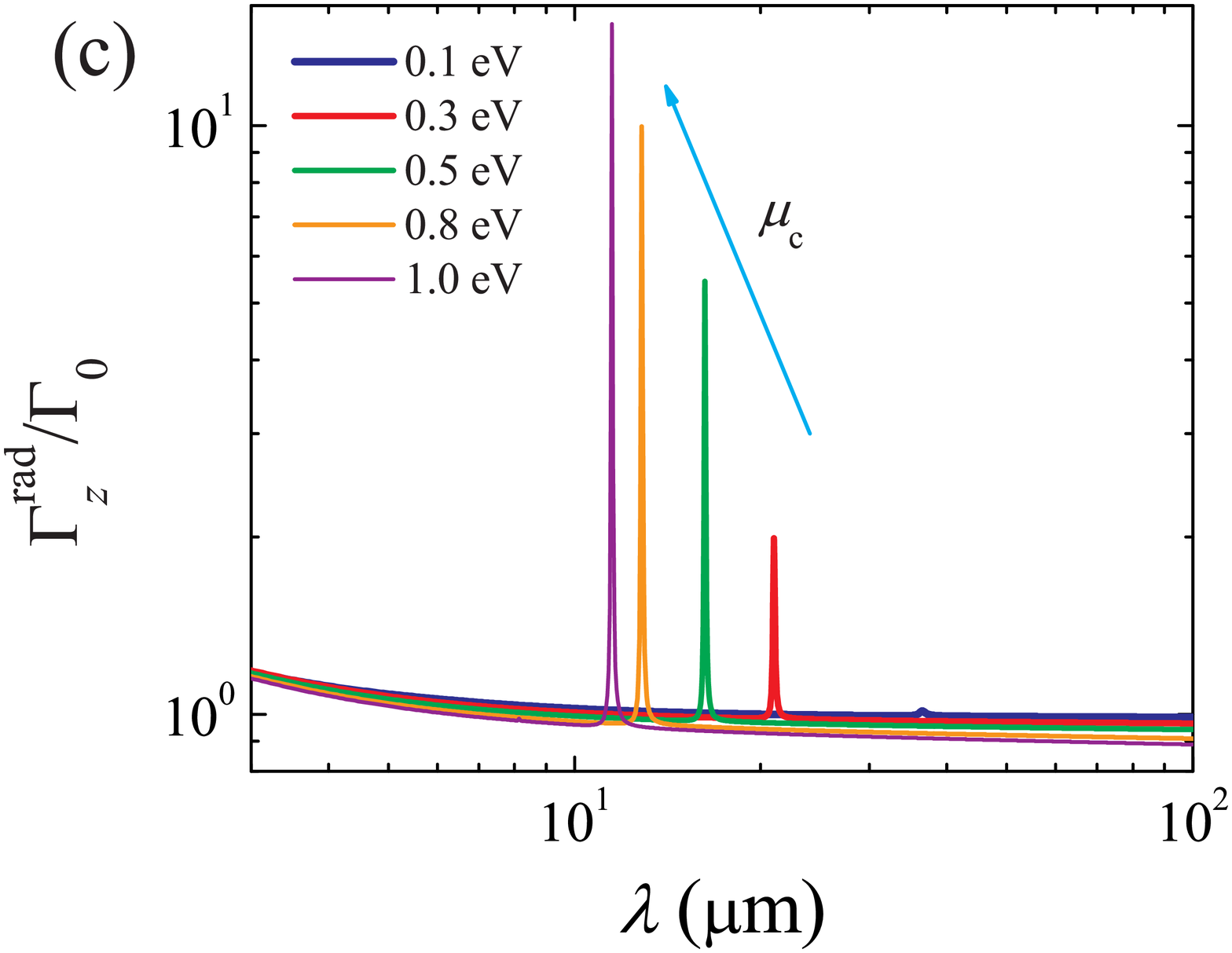}
\caption{Radiative decay rates for a dipole emitter near a graphene-coated dielectric nanowire of radius $a=100$~nm at $T=300$~K.
The distance between emitter and cylinder is $\Delta r=r'-b=10$~nm.
The plots show the influence of the graphene chemical potential $\mu_{\rm c}$ on the Fano line shapes in (a) $\Gamma_r^{\rm rad}$ and (b) $\Gamma_{\varphi}^{\rm rad}$, and on the Lorentzian line shape in (c) $\Gamma_z^{\rm rad}$ at the vicinity of the localized surface plasmon resonance.
The arrows indicate the direction of an increasing $\mu_{\rm c}$.
}\label{fig7}
\end{figure}

Let us consider the approximation in Eq.~(\ref{Qsca-TE-app}), where we have assumed $|a_1^{\rm TE}|^2\propto [(\omega^2-\omega_-^2)^2 + (\omega\gamma)^2]/[(\omega^2-\omega_+^2)^2 + (\omega\gamma)^2]$.
By using Eq.~(\ref{a1-TE}), after some algebra, we finally have
\begin{subequations}
\begin{align}
\frac{\Gamma_{r}^{\rm rad}(r',\omega)}{\Gamma_0} &\propto \frac{\left[X(\omega)+F_+(r')\right]^2}{[X(\omega)]^2+1},\label{Gamma-rad-r-app}\\
\frac{\Gamma_{\varphi}^{\rm rad}(r',\omega)}{\Gamma_0} &\propto\frac{\left[X(\omega)+F_-(r')\right]^2}{[X(\omega)]^2+1},\label{Gamma-rad-phi-app}
\end{align}
\end{subequations}
where we have defined
\begin{align}
X(\omega) &\equiv \displaystyle\frac{\omega-\omega_{+}}{\gamma/2},\\
F_{\pm}(r')&\equiv\pm\frac{b^2}{r'^2}\frac{(\omega_+-\omega_-)}{\gamma/2}.\label{Fano-para}
\end{align}
Since we are interested in the frequency range where the localized surface plasmon resonance occurs, we can consider that $\Gamma_z^{\rm rad}$ has a Lorentzian line shape as a function of $\omega$: ($\Gamma_{z}^{\rm rad}-\Gamma_0) \propto \Gamma_0/(X^2+1)$.
This Lorentzian line shape is related to the fact that $\Gamma_z^{\rm rad}$ depends only on the TM mode, see Eq.~(\ref{Gamma-cylinder-z-rad}).
In particular, the approximate prefactors to enter Eqs.~(\ref{Gamma-rad-r-app}) and (\ref{Gamma-rad-phi-app}) are $(1+b^2/r'^2)^2$ and $(1-b^2/r'^2)^2$, respectively (with $b<r'$).
For any dipole moment orientation, the maximum enhancement of $\Gamma_{\mathbf{d}_0}^{\rm rad}$ is achieved for $X(\omega_{\rm res})=0$, i.e., $\omega_{\rm res}=\omega_+$, which is defined in Eq.~(\ref{w-plus}).
The suppression of $\Gamma_r^{\rm rad}$ and $\Gamma_{\varphi}^{\rm rad}$ occurs when $X(\omega_{\rm inv})+F_{\pm}(r')=0$, which leads to
\begin{align}
{\omega}_{\rm inv}^{(\pm)}=\omega_+\pm\frac{b^2}{r'^2}(\omega_- - \omega_+),
\end{align}
where $\Gamma_{r}^{\rm rad}(\omega_{\rm inv}^{(+)})\ll\Gamma_0$ and $\Gamma_{\varphi}^{\rm rad}(\omega_{\rm inv}^{(-)})\ll\Gamma_0$.

In Fig.~\ref{fig6} we show that Eqs.~(\ref{Gamma-rad-r-app}) and (\ref{Gamma-rad-phi-app}) are good approximations for radiative decay rates at the frequency range of localized surface plasmon resonances.
The distance between point dipole and cylinder is $\Delta r=r'-b=80$~nm, which is still smaller than the radius of the dielectric core $a=100$~nm.
Note that we choose $\Delta r$ large enough to satisfy the approximations and small enough to achieve a strong enhancement of the radiative decay rate (see Fig.~\ref{fig4}).
Indeed, even for this value of $\Delta r$, we obtain $\Gamma_{\mathbf{d}_0}^{\rm rad}(\omega_{\rm res})\approx 10^3\Gamma_0$ associated with a large suppression $\Gamma_{\mathbf{d}_0}^{\rm rad}(\omega_{\rm inv})\approx 10^{-3}\Gamma_0$ within a spectral range of width $\Delta\lambda\approx 2.5~\mu$m.

It should be stressed that some deviations are expected in the case of imperfections and finite cylinders, mainly related to variations on guided-mode contributions, ohmic losses in the near field, and additional scattering by the edges at grazing angles.
However, for subwavelength-diameter cylinders, the effects of finite length are not expected to deteriorate the overall radiative contribution in the quasistatic limit, since the Fano effect induced in the radiative Purcell factor is based on an integral effect~\cite{Arruda_Springer219_2018}, i.e., it depends on the volume ratio of plasmonic coating and dielectric core.
Indeed, as discussed by Al\`u et al.~\cite{Alu_NJPhys12_2010}, the plasmonic cloaking of a thin dielectric infinite cylinder is not significantly affected by truncation effects.
Since we have unveiled the close relationship between Lorenz-Mie theory and the decay rates associated with cylindrical scatterers, it is expected that a similar conclusion can be applied to our analysis.

The Fano asymmetry parameter given in Eq.~(\ref{Fano-para}) depends explicitly on the distance $\Delta r$ between emitter and cylinder.
More importantly, we see that $F_{\pm}(r')\propto\sqrt{\mu_{\rm c}}$, which implies that both enhancement and suppression of the Purcell factor can be tuned by the chemical potential.
In practice the graphene chemical potential can be dynamically controlled by an applied static electric field (gate voltage) through the graphene/dielectric interface~\cite{Engheta_Sci332_2011,Basov_Nature487_2012}.
In the case of a graphene-coated nanowire, the experimental setup is similar to the one presented in Ref.~\cite{Basov_Nature487_2012}, with a bias electric field orthogonal to the graphene surface in order to change the plasmon frequency.

Without any approximation, we show in Fig.~\ref{fig7} the plots of radiative decay rates associated with a dipole emitter at $\Delta r=10$~nm from the graphene-coated nanowire.
By increasing the chemical potential from $\mu_{\rm c}=0.1$~eV to 1.0~eV, we show that the Fano line shape is blueshifted.
At the same time, the maximum enhancement of decay rates increases two orders of magnitude for $\Gamma_r^{\rm rad}(\lambda_{\rm res})$ and $\Gamma_{\varphi}^{\rm rad}(\lambda_{\rm res})$ [Figs.~\ref{fig7}(a) and \ref{fig7}(b), respectively] and one order of magnitude for $\Gamma_z^{\rm rad}(\lambda_{\rm res})$ [Fig.~\ref{fig7}(c)].
For a bias electric field orthogonal to the cylinder axis, one can neglect $\Gamma_z^{\rm rad}$ since the applied electric field will force the electric dipole moment $\mathbf{d}_0$ to orient in a direction parallel to it.
In particular, the suppression of radiative decay rates $\Gamma_r^{\rm rad}$ and $\Gamma_{\varphi}^{\rm rad}$ has opposite tendencies as a function of $\mu_{\rm c}$: $\Gamma_r^{\rm rad}(\lambda_{\rm inv})$ decreases whereas $\Gamma_{\varphi}^{\rm rad}(\lambda_{\rm inv})$ increases with increasing $\mu_{\rm c}$.
The small resonance peak that appears for high frequencies is due to the excitation of an electric quadrupole resonance $(\ell=2)$ within the plasmonic coating.
It is worth mentioning that our set of parameters does not correspond to an optimized configuration.
It is still possible to combine the variation on $\mu_{\rm c}$ and $r'$ with lower temperatures in order to achieve an even stronger enhancement or suppression of radiative decay rates~\cite{Cuevas_JQSRT200_2017}.

\section{Conclusion}
\label{conclusion}

Using the full-wave Lorenz-Mie solution, we have investigated the spontaneous-emission rate of a dipole emitter in the vicinity of a graphene-coated nanowire.
We have derived exact expressions for the radiative and nonradiative decay rates for three basic orientations of the dipole moment in relation to the cylinder.
Such analytical expressions can be straightforwardly generalized to circular multilayered cylinders in the framework of the Lorenz-Mie theory.
In the long wavelength limit, we have calculated approximate expressions for the radiative decay rate as a function of effective permittivities associated with the core-shell nanobody.
We have explicitly shown the connection between plasmonic Fano resonances in light scattering and
spontaneous emission of light.
More importantly, the enhancement and suppression of the radiative decay rate of a point dipole emitter near a graphene monolayer can be tuned by the graphene chemical potential monitored by a gate voltage.
The Fano asymmetry parameter of radiative decay rates, which determines the degree of asymmetry of the Fano line shape, is shown to be proportional to the square root of the chemical potential and depends strongly on the distance between dipole emitter and cylinder for a dipole moment oriented along $\varphi$ direction.
For a dipole moment oriented along $z$ direction, the interaction between dipole emitter and graphene is weak, leading to a Lorentzian line shape in the radiative decay rate.
The strong dependence of decay rates on the graphene chemical potential can be explored to enhance or suppress the radiative decay response of a plasmonic system by dynamically controlling the Fano resonance via a gate voltage.
As a result, the possibility of tuning and tailoring the spontaneous-emission rate near a graphene-based metamaterial that exhibits an asymmetric Fano line shape could lead to the engineering of low-loss nanophotonic devices for tunable single-photon sources.
In particular, these results could be explored in applications involving graphene coatings to achieve ultrahigh-contrast switching for spontaneous emission in specifically designed tunable plasmonic nanostructures~\cite{Gu_SciRep8_2018}.

\section*{Acknowledgments}
The authors acknowledge the Brazilian agencies for financial support.
T.J.A., R.B., and Ph.W.C. hold Grants from S\~ao Paulo Research Foundation (FAPESP) (Grants No. 2015/21194-3, No. 2014/01491-0, and No. 2013/04162-5, respectively).

\appendix

\section{Lorenz-Mie coefficients for core-shell cylinders}
\label{Lorenz-Mie}

The Lorenz-Mie coefficients associated with a center-symmetric core-shell cylinder are calculated from boundary conditions, reading~\cite{Arruda_JOpSocAmA31_2014}
\begin{align}
    a_{\ell}^{\rm TM}&=\frac{{\imath}}{\mu_2}\frac{\eta_2 J_{\ell}(\eta_2y)}{\eta_0 H_{\ell}^{(1)}(\eta_0y)}\frac{\left[\overline{\cal{V}}_{\ell}{\cal{P}}_{\ell}^{\rm TM}+\overline{\cal{C}}_{\ell}{\cal{Q}}_{\ell}^{\rm TM}\right]\gamma_{\ell}}{\overline{\cal{V}}_{\ell}{\cal{R}}_{\ell}^{\rm TM}},\label{an-TM}\\
    b_{\ell}^{\rm TM}&=\frac{1}{m_2}\frac{\eta_2 J_{\ell}(\eta_2y)}{\eta_0 H_{\ell}^{(1)}(\eta_0y)}\frac{\left[\overline{\cal{A}}_{\ell}{\cal{P}}_{\ell}^{\rm TM}+{\cal{Q}}_{\ell}^{\rm TM}\right]}{\overline{\cal{V}}_{\ell}{\cal{R}}_{\ell}^{\rm TM}}+\frac{J_{\ell}(\eta_0y)}{H_{\ell}^{(1)}(\eta_0y)},\\
    a_{\ell}^{\rm TE}&=\frac{1}{\mu_2}\frac{\eta_2 J_{\ell}(\eta_2y)}{\eta_0 H_{\ell}^{(1)}(\eta_0y)}\frac{\left[\overline{\cal{B}}_{\ell}{\cal{P}}_{\ell}^{\rm TE}+{\cal{Q}}_{\ell}^{\rm TE}\right]}{\overline{\cal{W}}_{\ell}{\cal{R}}_{\ell}^{\rm TE}}+\frac{J_{\ell}(\eta_0y)}{H_{\ell}^{(1)}(\eta_0y)},\\
    b_{\ell}^{\rm TE}&=-\frac{{\imath}}{m_2}\frac{\eta_2 J_{\ell}(\eta_2y)}{\eta_0 H_{\ell}^{(1)}(\eta_0y)}\frac{\left[\overline{\cal{W}}_{\ell}{\cal{P}}_{\ell}^{\rm TE}+\overline{\cal{D}}_{\ell}{\cal{Q}}_{\ell}^{\rm TE}\right]\gamma_{\ell}}{\overline{\cal{W}}_{\ell}{\cal{R}}_{\ell}^{\rm TE}},\label{bn-TE}
\end{align}
where both $\gamma_{\ell}\equiv \ell\cos\zeta$ and $\eta_q\equiv\sqrt{m_q^2-\cos^2\zeta}$ carry the dependence on $\zeta$, which is the complement of the incidence angle, with $m_q=\sqrt{\varepsilon_q\mu_q/(\varepsilon_0\mu_0)}$ and $q=\{0,1,2\}$.
Note that $\eta_0=\sqrt{1-\cos^2\zeta}=\sin\zeta$.
The auxiliary functions are
\begin{align*}
    \mathcal{A}_{\ell} &= \frac{J_{\ell}(\eta_2x)}{Y_{\ell}(\eta_2x)}\left[\frac{\mu_1L_{\ell}^{(J)}(\eta_1x)-\mu_2L_{\ell}^{(J)}(\eta_2x)}{\mu_1L_{\ell}^{(J)}(\eta_1x)-\mu_2L_{\ell}^{(Y)}(\eta_2x)}\right],\\
    \mathcal{B}_{\ell} &= \frac{J_{\ell}(\eta_2x)}{Y_{\ell}(\eta_2x)}\left[\frac{\varepsilon_1L_{\ell}^{(J)}(\eta_1x)-\varepsilon_2L_{\ell}^{(J)}(\eta_2x)}{\varepsilon_1L_{\ell}^{(J)}(\eta_1x)-\varepsilon_2L_{\ell}^{(Y)}(\eta_2x)}\right],\\
    \mathcal{C}_{\ell} &= \frac{\mu_2\left(1-\eta_2^2/\eta_1^2\right)}{{m}_2\eta_2^2x^2\left[\mu_1L_{\ell}^{(J)}(\eta_1x)-\mu_2L_{\ell}^{(Y)}(\eta_2x)\right]},\\
    \mathcal{D}_{\ell} &= \frac{\varepsilon_2\left(1-\eta_2^2/\eta_1^2\right)}{{m}_2\eta_2^2x^2\left[\varepsilon_1L_{\ell}^{(J)}(\eta_1x)-\varepsilon_2L_{\ell}^{(Y)}(\eta_2x)\right]},\\
    \mathcal{V}_{\ell} &= \frac{J_{\ell}(\eta_2x)}{Y_{\ell}(\eta_2x)}\mathcal{C}_{\ell},\\
    \mathcal{W}_{\ell} &= \frac{J_{\ell}(\eta_2x)}{Y_{\ell}(\eta_2x)}\mathcal{D}_{\ell},
\end{align*}
\begin{align*}
    \overline{\mathcal{A}}_{\ell} &= \frac{J_{\ell}(\eta_2y)}{Y_{\ell}(\eta_2y)}\left[\frac{\mu_2L_{\ell}^{(J)}(\eta_2y)-\mu_0 L_{\ell}^{(H)}(\eta_0y)}{\mu_2L_{\ell}^{(Y)}(\eta_2y)-\mu_0L_{\ell}^{(H)}(\eta_0y)}\right],\\
    \overline{\mathcal{B}}_{\ell} &= \frac{J_{\ell}(\eta_2y)}{Y_{\ell}(\eta_2y)}\left[\frac{\varepsilon_2L_{\ell}^{(J)}(\eta_2y)-\varepsilon_0L_{\ell}^{(H)}(\eta_0y)}{\varepsilon_2L_{\ell}^{(Y)}(\eta_2y)-\varepsilon_0L_{\ell}^{(H)}(\eta_0y)}\right],\\
    \overline{\mathcal{C}}_{\ell} &= \frac{\mu_2\left(1-\eta_0^2/\eta_2^2\right)}{{m}_2\eta_0^2y^2\left[\mu_2L_{\ell}^{(Y)}(\eta_2y)-\mu_0L_{\ell}^{(H)}(\eta_0y)\right]},\\
    \overline{\mathcal{D}}_{\ell} &= \frac{\varepsilon_2\left(1-\eta_0^2/\eta_2^2\right)}{{m}_2\eta_0^2y^2\left[\varepsilon_2L_{\ell}^{(Y)}(\eta_2y)-\varepsilon_0L_{\ell}^{(H)}(\eta_0y)\right]},\\
    \overline{\mathcal{V}}_{\ell} &= \frac{J_{\ell}(\eta_2y)}{Y_{\ell}(\eta_2y)}\overline{\mathcal{C}}_{\ell},\\
    \overline{\mathcal{W}}_{\ell} &= \frac{J_{\ell}(\eta_2y)}{Y_{\ell}(\eta_2y)}\overline{\mathcal{D}}_{\ell},
\end{align*}
where $x=ka$ and $y=kb$, and we have defined the logarithmic derivative function $L_{\ell}^{(Z)}(\rho)={Z_{\ell}'(\rho)}/[{\rho Z_{\ell}(\rho)}]$, with $Z_{\ell}$ being any special cylindrical function.
The remaining functions are
\begin{align*}
    {{\cal{F}}_{\ell}^{\rm  TM}}&=\frac{2{\imath} \varepsilon_2\left[\pi Y_{\ell}(\eta_2y)H_{\ell}^{(1)}(\eta_0y)\right]^{-1}}{\widetilde{m}_2\eta_2\eta_0y^2 \left[\varepsilon_0L_{\ell}^{(H)}(\eta_0y)-\varepsilon_2L_{\ell}^{(Y)}(\eta_2y)\right]},\\
    \frac{{\cal{P}}_{\ell}^{\rm  TM}}{{\cal{F}}_{\ell}^{\rm TM}}&={\overline{\cal{V}}}_{\ell}(1-\gamma_{\ell}^2{\cal{C}}_{\ell}{\cal{D}}_{\ell})-{\cal{V}}_{\ell}(1-\gamma_{\ell}^2\overline{\cal{C}}_{\ell}{{\cal{D}}}_{\ell})\nonumber\\
                                                      & +{\cal{B}}_{\ell}({\cal{C}}_{\ell}-{\overline{\cal{C}}}_{\ell}),\\
    \frac{{\cal{Q}}_{\ell}^{\rm TM}}{{\cal{F}}_{\ell}^{\rm TM}}&=\overline{\cal{{A}}}_{\ell}({\cal{V}}_{\ell}-{\cal{B}}_{\ell}{\cal{C}}_{\ell})-{\cal{A}}_{\ell}(\overline{\cal{V}}_{\ell}-{\cal{B}}_{\ell}\overline{\cal{C}}_{\ell})\nonumber\\
                                                      &+\gamma_{\ell}^2{\cal{W}}_{\ell}({\cal{C}}_{\ell}\overline{\cal{V}}_{\ell}-\overline{\cal{C}}_{\ell}{\cal{V}}_{\ell}),
    \end{align*}
        \begin{align*}
    {\cal{R}}_{\ell}^{\rm TM}&=({\cal{A}}_{\ell}-\gamma_{\ell}^2{\cal{C}}_{\ell}{\cal{W}}_{\ell})\Big[\gamma_{\ell}^2\overline{\cal{V}}_{\ell}({\cal{D}}_{\ell}-\overline{\cal{D}}_{\ell})\nonumber\\
                        &+ \overline{\cal{B}}_{\ell}(1-\gamma_{\ell}^2\overline{\cal{C}}_{\ell}{\cal{D}}_{\ell})-{\cal{B}}_{\ell}(1-\gamma_{\ell}^2\overline{\cal{C}}_{\ell}\overline{\cal{D}}_{\ell})\Big]\nonumber\\
                        & +(1-\gamma_{\ell}^2{\cal{C}}_{\ell}{\cal{D}}_{\ell})\Big[\overline{\cal{A}}_{\ell}({\cal{B}}_{\ell}-\overline{\cal{B}}_{\ell})\nonumber\\
                        & + \gamma_{\ell}^2\overline{\cal{W}}_{\ell}(\overline{\cal{V}}_{\ell}-{\cal{B}}_{\ell}\overline{\cal{C}}_{\ell})-\gamma_{\ell}^2{\cal{W}}_{\ell}(\overline{\cal{V}}_{\ell}-\overline{\cal{B}}_{\ell}\overline{\cal{C}}_{\ell})\Big]\nonumber\\
                        & +\gamma_{\ell}^2({\cal{V}}_{\ell}-{\cal{B}}_{\ell}{\cal{C}}_{\ell})\Big[({\cal{W}}_{\ell}-\overline{\cal{W}}_{\ell})\nonumber\\
                        &+\overline{\cal{D}}_{\ell}(\overline{\cal{A}}_{\ell}-\gamma_{\ell}^2\overline{\cal{C}}_{\ell}{\cal{W}}_{\ell})-{\cal{D}}_{\ell}(\overline{\cal{A}}_{\ell}-\gamma_{\ell}^2\overline{\cal{C}}_{\ell}\overline{\cal{W}}_{\ell})\Big],
    \end{align*}
    \begin{align*}
    {{\cal{F}}_{\ell}^{\rm  TE}}&=\frac{2{\imath} \mu_2\left[\pi Y_{\ell}(\eta_2y)H_{\ell}^{(1)}(\eta_0y)\right]^{-1}}{\eta_2\eta_0y^2 \left[\mu_0L_{\ell}^{(H)}(\eta_0y)-\mu_2L_{\ell}^{(Y)}(\eta_2y)\right]},\\
    \frac{{\cal{P}}_{\ell}^{\rm  TE}}{{\cal{F}}_{\ell}^{\rm TE}}&={\overline{\cal{W}}}_{\ell}(1-\gamma_{\ell}^2{\cal{C}}_{\ell}{\cal{D}}_{\ell})-{\cal{W}}_{\ell}(1-\gamma_{\ell}^2{{\cal{C}}}_{\ell}\overline{\cal{D}}_{\ell})\nonumber\\
                                                      & +{\cal{A}}_{\ell}({\cal{D}}_{\ell}-{\overline{\cal{D}}}_{\ell}),\\
    \frac{{\cal{Q}}_{\ell}^{\rm TE}}{{\cal{F}}_{\ell}^{\rm TE}}&=\overline{\cal{{B}}}_{\ell}({\cal{W}}_{\ell}-{\cal{A}}_{\ell}{\cal{D}}_{\ell})-{\cal{B}}_{\ell}(\overline{\cal{W}}_{\ell}-{\cal{A}}_{\ell}\overline{\cal{D}}_{\ell})\nonumber\\
                                                      &+\gamma_{\ell}^2{\cal{V}}_{\ell}({\cal{D}}_{\ell}\overline{\cal{W}}_{\ell}-\overline{\cal{D}}_{\ell}{\cal{W}}_{\ell})\ ,
    \end{align*}
    \begin{align*}
   {\cal{R}}_{\ell}^{\rm TE}&=({\cal{B}}_{\ell}-\gamma_{\ell}^2{\cal{D}}_{\ell}{\cal{V}}_{\ell})\Big[\gamma_{\ell}^2\overline{\cal{W}}_{\ell}({\cal{C}}_{\ell}-\overline{\cal{C}}_{\ell})\nonumber\\
                        &+ \overline{\cal{A}}_{\ell}(1-\gamma_{\ell}^2{\cal{C}}_{\ell}\overline{\cal{D}}_{\ell})-{\cal{A}}_{\ell}(1-\gamma_{\ell}^2\overline{\cal{C}}_{\ell}\overline{\cal{D}}_{\ell})\Big]\nonumber\\
                        & +(1-\gamma_{\ell}^2{\cal{C}}_{\ell}{\cal{D}}_{\ell})\Big[\overline{\cal{B}}_{\ell}({\cal{A}}_{\ell}-\overline{\cal{A}}_{\ell})\nonumber\\
                        &+ \gamma_{\ell}^2\overline{\cal{V}}_{\ell}(\overline{\cal{W}}_{\ell}-{\cal{A}}_{\ell}\overline{\cal{D}}_{\ell})-\gamma_{\ell}^2{\cal{V}}_{\ell}(\overline{\cal{W}}_{\ell}-\overline{\cal{A}}_{\ell}\overline{\cal{D}}_{\ell})\Big]\nonumber\\
                        & +\gamma_{\ell}^2({\cal{W}}_{\ell}-{\cal{A}}_{\ell}{\cal{D}}_{\ell})\Big[({\cal{V}}_{\ell}-\overline{\cal{V}}_{\ell})\nonumber\\
                        &+\overline{\cal{C}}_{\ell}(\overline{\cal{B}}_{\ell}-\gamma_{\ell}^2\overline{\cal{D}}_{\ell}{\cal{V}}_{\ell})-{\cal{C}}_{\ell}(\overline{\cal{B}}_{\ell}-\gamma_{\ell}^2\overline{\cal{D}}_{\ell}\overline{\cal{V}}_{\ell})\Big].
\end{align*}
Although it is not obvious, one can demonstrate that
\begin{align}
a_{\ell}^{\rm TM}=-b_{\ell}^{\rm TE}.
\end{align}
In addition, these coefficients follow the parity relations
\begin{align}
a_{-\ell}^{\rm TM}=-a_{\ell}^{\rm TM} \ &,\quad b_{-\ell}^{\rm TM}=b_{\ell}^{\rm TM} ,\\
a_{-\ell}^{\rm TE}=a_{\ell}^{\rm TE} \ &,\quad b_{-\ell}^{\rm TE}=-b_{\ell}^{\rm TE} .
\end{align}

For a coated cylinder normally irradiated with plane waves ($\zeta=90^{\rm o}$), Eqs.~(\ref{an-TM})--(\ref{bn-TE}) retrieve the well-known Lorenz-Mie coefficients for coated cylinders~\cite{Bohren_Book_1983}.
Indeed, the expressions for arbitrary $\zeta$ are simplified, leading to $a_{\ell}^{\rm TM}|_{\zeta=90^{\rm o}}=b_{\ell}^{\rm TE}|_{\zeta=90^{\rm o}}=0$ and
\begin{align}
    a_{\ell}^{\rm TE}\big|_{\zeta=90^{\rm o}}&=\frac{\widetilde{m}_2J_{\ell}'(y)\alpha_{\ell} -J_{\ell}(y)}
        {\widetilde{m}_2H_{\ell}'^{(1)}(y)\alpha_{\ell}-H_{\ell}^{(1)}(y)},\\
    b_{\ell}^{\rm TM}\big|_{\zeta=90^{\rm o}}&=\frac{J_{\ell}'(y)\beta_{\ell}-\widetilde{m}_2J_{\ell}(y)}
         {H_{\ell}'^{(1)}(y)\beta_{\ell}-\widetilde{m}_2H_{\ell}^{(1)}(y)},
\end{align}
with new auxiliary functions
\begin{align*}
\alpha_{\ell}&=\frac{J_{\ell}(m_2y)-{\cal{A}}_{\ell}Y_{\ell}(m_2y)}{J_{\ell}'(m_2y)-{\cal{A}}_{\ell}Y_{\ell}'(m_2y)},\\
\beta_{\ell}&=\frac{J_{\ell}(m_2y)-{\cal{B}}_{\ell}Y_{\ell}(m_2y)}{J_{\ell}'(m_2y)-{\cal{B}}_{\ell}Y_{\ell}'(m_2y)},\\
    {\cal{A}}_{\ell}&=\frac{\widetilde{m}_1J_{\ell}(m_1x)J_{\ell}'(m_2x)-\widetilde{m}_2J_{\ell}'(m_1x)J_{\ell}(m_2x)}{\widetilde{m}_1J_{\ell}(m_1x)Y_{\ell}'(m_2x)-\widetilde{m}_2J_{\ell}'(m_1x)Y_{\ell}(m_2x)},\\
    {\cal{B}}_{\ell}&=\frac{\widetilde{m}_2J_{\ell}(m_1x)J_{\ell}'(m_2x)-\widetilde{m}_1J_{\ell}'(m_1x)J_{\ell}(m_2x)}{\widetilde{m}_2J_{\ell}(m_1x)Y_{\ell}'(m_2x)-\widetilde{m}_1J_{\ell}'(m_1x)Y_{\ell}(m_2x)},
\end{align*}
where $\widetilde{m}_q=\sqrt{\varepsilon_q\mu_0/(\varepsilon_0\mu_q)}$.
In particular, the case of light scattering by a homogeneous cylinder $(\varepsilon_1,\mu_1)$ of radius $b$ can be readily obtained from the expressions above by imposing $(\varepsilon_1,\mu_1)=(\varepsilon_2,\mu_2)$, i.e., $\mathcal{A}_{\ell}=\mathcal{B}_{\ell}=0$~\cite{Bohren_Book_1983}.

\section{Radiative and nonradiative decay rates of a dipole emitter in the vicinity of a cylinder}
\label{decay-rates}

Let us calculate the radiative decay rate associated with the system described in Sec.~\ref{decay} by using the full-wave Lorenz-Mie theory.
Using Eq.~(\ref{definition}) and recalling that $\int_0^{2\pi}{\rm d}\varphi e^{\imath(\ell-\ell')\varphi}=2\pi\delta_{\ell\ell'}$, where $\delta_{\ell\ell'}$ is the Kronecker delta, we obtain for the three basic dipole moment orientations $\mathbf{d}_0=d_0\hat{\mathbf{r}}$, $d_0\hat{\boldsymbol{\varphi}}$ and $d_0\hat{\mathbf{z}}$, respectively:
\begin{widetext}
\begin{subequations}
\begin{align}
\frac{\Gamma_r^{\rm rad}(kr')}{\Gamma_{0}}
&=\frac{3}{2}\sum_{\ell=-\infty}^{\infty}\int_0^{\pi/2}{\rm d}\zeta\sin\zeta
\Bigg\{\left|\ell\frac{a_{\ell}^{\rm TM} H_{\ell}^{(1)}(kr'\sin\zeta)}{kr'\sin\zeta} +\imath \cos\zeta\left(b_{\ell}^{\rm TM} H_{\ell}'^{(1)}(kr'\sin\zeta)-J_{\ell}'(kr'\sin\zeta)\right)\right|^2\nonumber\\
&+\left|\ell\left(\frac{a_{\ell}^{\rm TE} H_{\ell}^{(1)}(kr'\sin\zeta)-J_{\ell}(kr'\sin\zeta)}{kr'\sin\zeta}\right) + \imath \cos\zeta b_{\ell}^{\rm TE} H_{\ell}'^{(1)}(kr'\sin\zeta)\right|^2 \Bigg\},\label{Gamma-cylinder-r-rad}\\
\frac{\Gamma_{\varphi}^{\rm rad}(kr')}{\Gamma_{0}}
&=\frac{3}{2}\sum_{\ell=-\infty}^{\infty}\int_0^{\pi/2}{\rm d}\zeta\sin\zeta
\Bigg\{\left|\ell\cos\zeta\frac{b_{\ell}^{\rm TE} H_{\ell}^{(1)}(kr'\sin\zeta)}{kr'\sin\zeta} -\imath \left(a_{\ell}^{\rm TE} H_{\ell}'^{(1)}(kr'\sin\zeta)-J_{\ell}'(kr'\sin\zeta)\right)\right|^2\nonumber\\
&+\left|\ell\cos\zeta\left(\frac{b_{\ell}^{\rm TM} H_{\ell}^{(1)}(kr'\sin\zeta)-J_{\ell}(kr'\sin\zeta)}{kr'\sin\zeta}\right) - \imath  a_{\ell}^{\rm TM} H_{\ell}'^{(1)}(kr'\sin\zeta)\right|^2 \Bigg\},\label{Gamma-cylinder-phi-rad}\\
\frac{\Gamma_{z}^{\rm rad}(kr')}{\Gamma_{0}}&=\frac{3}{2}\sum_{\ell=-\infty}^{\infty}\int_0^{\pi/2}{\rm d}\zeta\sin^3\zeta\left[\left|a_{\ell}^{\rm TM} H_{\ell}^{(1)}(kr'\sin\zeta)\right|^2+\left|b_{\ell}^{\rm TM} H_{\ell}^{(1)}(kr'\sin\zeta)-J_{\ell}(kr'\sin\zeta)\right|^2\right],\label{Gamma-cylinder-z-rad}
\end{align}
\end{subequations}
\end{widetext}
where we have considered Eqs.~(\ref{Einc-TM})--(\ref{Esca-TM}), taking into account both TM and TE modes: $\mathbf{E}_{\rm vac}=\mathbf{E}_{\rm vac}^{\rm TM}+\mathbf{E}_{\rm vac}^{\rm TE}$ and $\mathbf{E}_{\rm sca}=\mathbf{E}_{\rm sca}^{\rm TM}+\mathbf{E}_{\rm sca}^{\rm TE}$.
The radiative decay rate of a dipole moment $\mathbf{d}_0$ randomly oriented in relation to the cylindrical surface is simply the spatial mean: $\Gamma^{\rm rad}=(\Gamma_r^{\rm rad} + \Gamma_{\varphi}^{\rm rad}+\Gamma_z^{\rm rad})/3$.
Note that for the $z$ direction, only the TM mode contributes to the radiative decay rate.
Of course, in the absence of cylinder, we have $a_{\ell}^{\rm TM}=a_{\ell}^{\rm TE}=b_{\ell}^{\rm TM}=b_{\ell}^{\rm TE}=0$ and hence $\Gamma_r^{\rm rad}=\Gamma_{\varphi}^{\rm rad}=\Gamma_z^{\rm rad}=\Gamma_0$.

It is worth emphasizing that $a_{\ell}^{\rm TM}$, $a_{\ell}^{\rm TE}$, $b_{\ell}^{\rm TM}$, and $b_{\ell}^{\rm TE}$ are the usual Lorenz-Mie coefficients associated with a cylindrical scatterer under oblique incidence of plane waves~\cite{Bohren_Book_1983,Wait_CanJPhys33_1955}.
As a result, one can easily generalize the present calculations to multilayered cylinders by choosing properly the classical scattering coefficients $a_{\ell}$ and $b_{\ell}$ to enter into Eqs.~(\ref{Gamma-cylinder-r-rad})--(\ref{Gamma-cylinder-z-rad})~\cite{Kleiman_JQSRT63_1999}.
Here, we are interested in a single-layered core-shell cylinder according to Fig.~\ref{fig1} and with scattering coefficients provided in Appendix~\ref{Lorenz-Mie}.

The total decay rate, which takes into account radiative and nonradiative contributions, can be written as $\Gamma_{\mathbf{d}_0}^{\rm total}/\Gamma_0=1 + 6\pi\varepsilon_0 {\rm Im}[\mathbf{d}_0\cdot\mathbf{E}_{\mathbf{d}_0}^{\rm sca}(\mathbf{r}')]/d_0^2k^3$, where $\mathbf{E}_{\mathbf{d}_0}^{\rm sca}(\mathbf{r}')$ is the scattering part of electric field associated with the dipole source~\cite{Carminati_SurfSciRep70_2015,Belov_SciRep5_2015}.
As a result, the corresponding frequency shift $\delta\omega_{\mathbf{d}_0}$ in the transition frequency due to the presence of a nanobody is $\delta\omega_{\mathbf{d}_0}/\Gamma_0=3\pi\varepsilon_0 {\rm Re}[\mathbf{d}_0\cdot\mathbf{E}_{\mathbf{d}_0}^{\rm sca}(\mathbf{r}')]/d_0^2k^3$~\cite{Letokhov_JModOpt43_2_1996,Carminati_SurfSciRep70_2015}.
Finding an analytical expression for $\mathbf{E}_{\mathbf{d}_0}^{\rm sca}(\mathbf{r}')$ is in general complicated~\cite{Chew_JCPhys87_1987}.
However, once we have $\Gamma_{\mathbf{d}_0}^{\rm rad}$ as described in Eqs.~(\ref{Gamma-cylinder-r-rad})--(\ref{Gamma-cylinder-z-rad}), we can calculate $\Gamma_{\mathbf{d}_0}^{\rm total}$ indirectly by using the energy conservation in the Lorenz-Mie theory~\cite{Chew_JCPhys87_1987,Dujardin_OptExp16_2008,Arruda_PhysRevA96_2017}.
Indeed, we recall that the scattering, extinction and absorption efficiencies (or normalized cross sections) of a cylindrical scatterer are~\cite{Bohren_Book_1983}
\begin{subequations}
\begin{align}
Q_{\rm sca}^{\rm TM}&=\frac{2}{kb}\sum_{\ell=-\infty}^{\infty}\left(\left|a_{\ell}^{\rm TM}\right|^2+\left|b_{\ell}^{\rm TM}\right|^2\right),\label{Qsca}\\
Q_{\rm ext}^{\rm TM}&=\frac{2}{kb}\sum_{\ell=-\infty}^{\infty}{\rm Re}\left(b_{\ell}^{\rm TM}\right),\label{Qext}\\
Q_{\rm abs}^{\rm TM}&=Q_{\rm ext}^{\rm TM}-Q_{\rm sca}^{\rm TM},\label{Qabs}
\end{align}
\end{subequations}
respectively, where the corresponding $Q_{\rm sca}^{\rm TE}$, $Q_{\rm ext}^{\rm TE}$, and $Q_{\rm abs}^{\rm TE}$ are readily obtained by replacing $(a_{\ell}^{\rm TM},b_{\ell}^{\rm TM})$ with ($b_{\ell}^{\rm TE},a_{\ell}^{\rm TE}$).
For nondissipative media, one has ${\rm Re}(b_{\ell}^{\rm TM})=|a_{\ell}^{\rm TM}|^2+|b_{\ell}^{\rm TM}|^2$ and ${\rm Re}(a_{\ell}^{\rm TE})=|a_{\ell}^{\rm TE}|^2+|b_{\ell}^{\rm TE}|^2$.
As demonstrated in Ref.~\cite{Chew_JCPhys87_1987} for the spherical case, this simple observation allows us to calculate the total emission rate from the radiative emission rate in the Lorenz-Mie framework.
The main assumption is that the nonradiative contribution to the spontaneous-emission rate comes from ohmic losses on the surface of the nanobody.
Lets us consider, e.g., $\Gamma_z^{\rm rad}$.
Rewriting Eq.~(\ref{Gamma-cylinder-z-rad}) and using $\sum_{\ell=-\infty}^{\infty}J_{\ell}(\rho)^2=1$, we obtain
\begin{align}
\frac{\Gamma_{z}^{\rm rad}(kr')}{\Gamma_{0}}&=1+\frac{3}{2}\sum_{\ell=-\infty}^{\infty}\int_0^{\pi/2}{\rm d}\zeta\sin^3\zeta\nonumber\\
&\times\Bigg\{\left(\left|a_{\ell}^{\rm TM}\right|^2+\left|b_{\ell}^{\rm TM} \right|^2\right)\left|H_{\ell}^{(1)}(kr'\sin\zeta)\right|^2\nonumber\\
&-2{\rm Re}\left[b_{\ell}^{\rm TM} H_{\ell}^{(1)}(kr'\sin\zeta)J_{\ell}(kr'\sin\zeta)\right]\Bigg\}.\label{Gamma-cylinder-z-rad2}
\end{align}
The total decay rate associated with the $z$ component is readily obtained from Eq.~(\ref{Gamma-cylinder-z-rad2}) by replacing $|a_{\ell}^{\rm TM}|^2+|b_{\ell}^{\rm TM} |^2$ with ${\rm Re}(b_{\ell}^{\rm TM} )$.
Using the same idea for $r$ and $\varphi$ components, after some algebra, we finally have
\begin{widetext}
\begin{subequations}
\begin{align}
\frac{\Gamma_r^{\rm total}(kr')}{\Gamma_{0}}
&=1-\frac{3}{2}{\rm Re}\sum_{\ell=-\infty}^{\infty}\int_0^{\pi/2}{\rm d}\zeta\sin\zeta
\Bigg\{
{\ell^2}a_{\ell}^{\rm TE} \left[\frac{H_{\ell}^{(1)}(kr'\sin\zeta)}{kr'\sin\zeta}\right]^2
+\cos^2\zeta b_{\ell}^{\rm TM} \left[H_{\ell}'^{(1)}(kr'\sin\zeta)\right]^2\nonumber\\
&-2\imath\ell\cos\zeta a_{\ell}^{\rm TM*}\frac{H_{\ell}^{(1)*}(kr'\sin\zeta)}{kr'\sin\zeta}\left[b_{\ell}^{\rm TM}H_{\ell}'^{(1)}(kr'\sin\zeta)-J_{\ell}'(kr'\sin\zeta)\right]\nonumber\\
&+2\imath\ell\cos\zeta b_{\ell}^{\rm TE*}{H_{\ell}'^{(1)*}(kr'\sin\zeta)}\left[\frac{a_{\ell}^{\rm TE}H_{\ell}^{(1)}(kr'\sin\zeta)-J_{\ell}(kr'\sin\zeta)}{kr'\sin\zeta}\right] \Bigg\},\label{Gamma-cylinder-r-total}\\
\frac{\Gamma_{\varphi}^{\rm total}(kr')}{\Gamma_{0}}&=1-\frac{3}{2}{\rm Re}\sum_{\ell=-\infty}^{\infty}\int_0^{\pi/2}{\rm d}\zeta\sin\zeta\Bigg\{
a_{\ell}^{\rm TE} \left[H_{\ell}'^{(1)}(kr'\sin\zeta)\right]^2+\ell^2\cos^2\zeta b_{\ell}^{\rm TM} \left[\frac{H_{\ell}^{(1)}(kr'\sin\zeta)}{kr'\sin\zeta}\right]^2\nonumber\\
&-2\imath\ell\cos\zeta a_{\ell}^{\rm TM*}{H_{\ell}'^{(1)*}(kr'\sin\zeta)}\left[\frac{b_{\ell}^{\rm TM}H_{\ell}^{(1)}(kr'\sin\zeta)-J_{\ell}(kr'\sin\zeta)}{kr'\sin\zeta}\right]\nonumber\\
&+2\imath\ell\cos\zeta b_{\ell}^{\rm TE*}\frac{H_{\ell}^{(1)*}(kr'\sin\zeta)}{kr'\sin\zeta}\left[a_{\ell}^{\rm TE}H_{\ell}'^{(1)}(kr'\sin\zeta)-J_{\ell}'(kr'\sin\zeta)\right] \Bigg\},\label{Gamma-cylinder-phi-total}\\
\frac{\Gamma_{z}^{\rm total}(kr')}{\Gamma_{0}}&=1-\frac{3}{2}{\rm Re}\sum_{\ell=-\infty}^{\infty}\int_0^{\pi/2}{\rm d}\zeta\sin^3\zeta b_{\ell}^{\rm TM} \left[H_{\ell}^{(1)}(kr'\sin\zeta)\right]^2,\label{Gamma-cylinder-z-total}
\end{align}
\end{subequations}
\end{widetext}
where we have used the sums~\cite{Klimov_PhysRevA69_2004} $\sum_{\ell=-\infty}^{\infty} [\ell J_{\ell}(\rho)/\rho]^2=1/2$ and $\sum_{\ell=-\infty}^{\infty}J_{\ell}'(\rho)^2=1/2$.
Once again, for a dipole moment with arbitrary orientation in relation to the cylindrical surface, one has the spatial mean $\Gamma^{\rm total}=(\Gamma_r^{\rm total} + \Gamma_{\varphi}^{\rm total}+\Gamma_z^{\rm total})/3$.
Subtracting Eqs.~(\ref{Gamma-cylinder-r-rad})--(\ref{Gamma-cylinder-z-rad}) from the corresponding Eqs.~(\ref{Gamma-cylinder-r-total})--(\ref{Gamma-cylinder-z-total}), we calculate the nonradiative decay rates for each dipole moment orientation:
\begin{widetext}
\begin{subequations}
\begin{align}
\frac{\Gamma_r^{\rm nrad}(kr')}{\Gamma_{0}}&=\frac{3}{2}{\rm Re}\sum_{\ell=-\infty}^{\infty}\int_0^{\pi/2}{\rm d}\zeta\sin\zeta\Bigg\{\left(a_{\ell}^{\rm TE} -\left|a_{\ell}^{\rm TE} \right|^2-\left|b_{\ell}^{\rm TE} \right|^2\right)\ell^2\left|\frac{H_{\ell}^{(1)}(kr'\sin\zeta)}{kr'\sin\zeta}\right|^2\nonumber\\
&\qquad\qquad\qquad\qquad\qquad\quad+\left(b_{\ell}^{\rm TM} -\left|a_{\ell}^{\rm TM} \right|^2-\left|b_{\ell}^{\rm TM} \right|^2\right)\cos^2\zeta \left|H_{\ell}'^{(1)}(kr'\sin\zeta)\right|^2\Bigg\},\label{Gamma-cylinder-r-nrad}\\
\frac{\Gamma_{\varphi}^{\rm nrad}(kr')}{\Gamma_{0}}&=\frac{3}{2}{\rm Re}\sum_{\ell=-\infty}^{\infty}\int_0^{\pi/2}{\rm d}\zeta\sin\zeta\Bigg\{\left(a_{\ell}^{\rm TE} -\left|a_{\ell}^{\rm TE} \right|^2-\left|b_{\ell}^{\rm TE} \right|^2\right)\left|H_{\ell}'^{(1)}(kr'\sin\zeta)\right|^2\nonumber\\
&\qquad\qquad\qquad\qquad\qquad\quad+\left(b_{\ell}^{\rm TM} -\left|a_{\ell}^{\rm TM} \right|^2-\left|b_{\ell}^{\rm TM} \right|^2\right)\ell^2\cos^2\zeta\left|\frac{H_{\ell}^{(1)}(kr'\sin\zeta)}{kr'\sin\zeta}\right|^2\Bigg\},\\
\frac{\Gamma_{z}^{\rm nrad}(kr')}{\Gamma_{0}}&=\frac{3}{2}{\rm Re}\sum_{\ell=-\infty}^{\infty}\int_0^{\pi/2}{\rm d}\zeta\sin^3\zeta\left\{\left(b_{\ell}^{\rm TM} -\left|a_{\ell}^{\rm TM} \right|^2-\left|b_{\ell}^{\rm TM} \right|^2\right)\left|H_{\ell}^{(1)}(kr'\sin\zeta)\right|^2\right\}.\label{Gamma-cylinder-z-nrad}
\end{align}
\end{subequations}
\end{widetext}
The corresponding frequency shifts $\delta\omega_{r}$, $\delta\omega_{\varphi}$, and $\delta\omega_{z}$ due to the presence of the cylinder are obtained from Eqs.~(\ref{Gamma-cylinder-r-total})--(\ref{Gamma-cylinder-z-total}), respectively, by replacing $(\Gamma_{\mathbf{d}_0}^{\rm total}-\Gamma_0)$ with $2\delta\omega_{\mathbf{d}_0}$ and ${\rm Re}(\ldots)$ with $-{\rm Im}(\ldots)$.

Equations~(\ref{Gamma-cylinder-r-rad})--(\ref{Gamma-cylinder-z-rad}) and Eqs.~(\ref{Gamma-cylinder-r-total})--(\ref{Gamma-cylinder-z-nrad}) are the main analytical result of this paper.
 As a limiting case of these expressions, one can use Refs.~\cite{Klimov_PhysRevA62_2000,Klimov_PhysRevA69_2004}, where a different approach was applied to calculate the decay rates related to a dipole emitter on the surface of a homogeneous dielectric cylinder.
We have verified that the expressions above reproduce all the plots in Ref.~\cite{Klimov_PhysRevA69_2004} for $r'=b$, $\varepsilon_1=\varepsilon_2$, and $\mu_1=\mu_2=\mu_0$.
In general, by properly defining the electric Green's tensor of the system one can straightforwardly derive the Purcell factor via the LDOS~\cite{Carminati_SurfSciRep70_2015}.
Indeed, several approaches are already available to calculate the Purcell effect regarding a point-dipole emitter in cylindrical geometry using the standard definition of LDOS~\cite{Bradley_PhysRevA89_2014} or mode decomposition of the electromagnetic field~\cite{Zakowicz_PhysRevA62_2000}.
Notwithstanding the available studies, Eqs.~(\ref{Gamma-cylinder-r-rad})--(\ref{Gamma-cylinder-z-rad}) and Eqs.~(\ref{Gamma-cylinder-r-total})--(\ref{Gamma-cylinder-z-nrad}) are original and seem to be the most natural choice for the cylindrical geometry owing to the explicit connection between decay rates and Lorenz-Mie theory.
For the spherical geometry, this connection is well known and widely explored in both classical and quantum-mechanical approaches~\cite{Chew_JCPhys87_1987}.


\begin{thebibliography}{99}



\bibitem{Purcell_PhysRev69_1946}
E. M. Purcell,
Phys. Rev. {\bf 69}, 681 (1946).

\bibitem{Chew_JCPhys87_1987}
H. Chew,
J. Chem. Phys. {\bf 87}, 1355 (1987).

\bibitem{Dereux_PhysRevB84_2011}
J. Barthes, G. Colas des Francs, A. Bouhelier, J. C. Weeber, and A. Dereux.
Phys. Rev. B {\bf 84}, 073403 (2011).

 \bibitem{Bordo_JOSAB31_2014}
K. V. Filonenko, M. Willatzen, and V. G. Bordo,
J. Opt. Soc. Am. B {\bf 31}, 2002 (2014).

\bibitem{Belov_SciRep5_2015}
A. E. Krasnok, A. P. Slobozhanyuk, C. R. Simovski, S. A. Tretyakov, A. N. Poddubny, A. E. Miroshnichenko, Y. S. Kivshar, and P. A. Belov,
Sci. Rep. {\bf 5}, 12956 (2015).

\bibitem{Carminati_SurfSciRep70_2015}
 R. Carminati, A. Caz\'e, D. Cao, F. Peragut, V. Krachmalnicoff, R. Pierrat, and Y. De Wilde,
Surf. Sci. Rep. {\bf 70}, 1 (2015).

\bibitem{Sandoghdar_Nature405_2000}
J. Michaelis, C. Hettich, J. Mlynek, and V. Sandoghdar,
Nature {\bf 405}, 325 (2000).

\bibitem{Lukin_NatPhys3_2007}
D. E. Chang, A. S. S{\o}rensen, E. A. Demler, and M. D. Lukin,
Nat. Phys. {\bf 3}, 807 (2007).

\bibitem{Gallego_PhysRevLett121_2018}
J. Gallego, W. Alt, T. Macha, M. Martinez-Dorantes, D. Pandey, and D. Meschede,
Phys. Rev. Lett. {\bf 121}, 173603 (2018).

\bibitem{Berini_PhysRevB78_2008}
I. De Leon and P. Berini,
Phys. Rev. B {\bf 78}, 161401(R) (2008).

\bibitem{Fainman_OptExp21_2013}
Q. Gu, B. Slutsky, F. Vallini, J. S. T. Smalley,
 M. P. Nezhad, N. C. Frateschi, and Y. Fainman,
Opt. Express {\bf 21}, 15603 (2013).

\bibitem{Girard_JOpt18_2016}
G. Colas des Francs, J. Barthes, A. Bouhelier, J. C. Weeber, A. Dereux, A. Cuche, and C. Girard,
J. Opt. {\bf 18}, 094005 (2016).

\bibitem{Arruda_PhysRevA96_2017}
T. J. Arruda, R. Bachelard, J. Weiner, S. Slama, and P. W. Courteille,
Phys. Rev. A {\bf 96}, 043869 (2017).

\bibitem{Cuevas_JOpt18_2016}
M. Cuevas,
J. Opt. {\bf 18}, 105003 (2016).

\bibitem{Farina_PhysRevA87_2013}
W. J. M. Kort-Kamp, F. S. S. Rosa, F. A. Pinheiro, and C. Farina,
Phys. Rev. A {\bf 87}, 023837 (2013).

\bibitem{Liu_Nature9_2014}
D. Lu, J. J. Kan, E. E. Fullerton, and  Z. Liu,
Nat. Nanotechnol. {\bf 9}, 48 (2014).

\bibitem{Szilard_PhysRevB94_2016}
D. Szilard, W. J. M. Kort-Kamp, F. S. S. Rosa, F. A. Pinheiro, and C. Farina,
Phys. Rev. B {\bf 94}, 134204 (2016).

\bibitem{Lukin_PhysRevLett97_2006}
D. E. Chang, A. S. S{\o}rensen, P. R. Hemmer, and M. D. Lukin,
Phys. Rev. Lett. {\bf 97}, 053002 (2006).

\bibitem{Lukin_Nature450_2007}
A. V. Akimov, A. Mukherjee, C. L. Yu, D. E. Chang, A. S. Zibrov, P. R. Hemmer, H. Park, and M. D. Lukin,
Nature {\bf 450}, 402 (2007).

\bibitem{Sun_SciApp4_2015}
Y. Fang and M. Sun,
Light Sci. Appl {\bf 4}, e294 (2015).

\bibitem{Gu_SciRep8_2018}
H. Hao, J. Ren, X. Duan, G. Lu, I. C. Khoo, Q. Gong, and Y. Gu,
Sci. Rep. {\bf 8}, 11244 (2018).

\bibitem{Philpott_JChemPhys62_1975}
M. R. Philpott,
J. Chem. Phys. {\bf 62}, 1812 (1975).

\bibitem{Eagen_OptLett4_1979}
W. H. Weber and C. F. Eagen,
Opt. Lett. {\bf 4}, 236 (1979).

\bibitem{Bohren_Book_1983}
C. F. Bohren and D. R. Huffman,
{\it Absorption and Scattering of Light by Small Particles}
(Wiley, New York, 1983).

\bibitem{Soljacic_PhysRevB80_2009}
M. Jablan, H. Buljan, and M. Soljacic,
Phys. Rev. B {\bf 80}, 245435 (2009).

\bibitem{Yakovlev_PhysRevB86_2012}
G. W. Hanson, E. Forati, W. Linz, and A. B. Yakovlev,
Phys. Rev. B {\bf 86}, 235440 (2012).

\bibitem{Alu_PhysRevB80_2009}
A. Al\`u,
Phys. Rev. B {\bf 80}, 245115 (2009).

\bibitem{Alu_ACSNano5_2011}
P.-Y. Chen and A. Al\`u,
ACS Nano {\bf 5}, 5855 (2011).

\bibitem{Abajo_NanoLett11_2011}
F. H. L. Koppens, D. E. Chang, and F. J. Garc\'ia de Abajo,
Nano Lett. {\bf 11}, 3370 (2011).

\bibitem{Wang_NatNano6_2011}
L. Ju, B. Geng, J. Horng, C. Girit, M. Martin, Z. Hao, H. A. Bechtel, X. Liang, A. Zettl, Y. R. Shen, and F. Wang,
Nat. Nanotechnol. {\bf 6}, 630 (2011).

\bibitem{Engheta_Sci332_2011}
 A. Vakil and N. Engheta,
 Science {\bf 332}, 1291 (2011).

 \bibitem{Basov_Nature487_2012}
Z. Fei, A. S. Rodin, G. O. Andreev, W. Bao, A. S. McLeod, M. Wagner, L. M. Zhang, Z. Zhao, M. Thiemens, G. Dominguez, M. M. Fogler, A. H. Castro Neto, C. N. Lau, F. Keilmann, and D. N. Basov,
Nature {\bf 487}, 82 (2012).

\bibitem{Cuevas_JOpt17_2015}
M. Riso, M. Cuevas, and R. A. Depine,
J. Opt. {\bf 17}, 075001 (2015).

\bibitem{Farina_PhysRevB92_2015}
W. J. M. Kort-Kamp, B. Amorim, G. Bastos, F. A. Pinheiro, F. S. S. Rosa, N. M. R. Peres, and C. Farina,
Phys. Rev. B {\bf 92}, 205415 (2015).

\bibitem{Cuevas_JQSRT173_2016}
M. Cuevas, M. A. Riso, and R. A. Depine,
J. Quant. Spectrosc. Radiat. Transf. {\bf 173} 26 (2016).

\bibitem{Arruda_JOpSocAmA31_2014}
T. J. Arruda, A. S. Martinez, and F. A. Pinheiro,
J. Opt. Soc. Am. A {\bf 31}, 1811 (2014).

\bibitem{Fano_PhysRev124_1961}
U. Fano,
Phys. Rev. {\bf 124}, 1866 (1961).

\bibitem{Kivshar_RevModPhys82_2010}
A. E. Miroshnichenko, S. Flach, and Y. S. Kivshar,
Rev. Mod. Phys. {\bf 82}, 2257 (2010).

\bibitem{Wu_SciRep7_2017}
S. Zhang, J. Li, R. Yu, W. Wang, and Y. Wu,
Sci. Rep. {\bf 7}, 39781 (2017).

\bibitem{Weis_Sci330_2010}
S. Weis, R. Rivi\`ere, S. Del\'eglise, E. Gavartin, O. Arcizet, A. Schliesser, and T. J. Kippenberg,
Science {\bf 330}, 1520 (2010).

\bibitem{Painter_Nature472_2011}
A. H. Safavi-Naeini, T. P. Mayer Alegre, J. Chan, M. Eichenfield, M. Winger, Q. Lin, J. T. Hill, D. E. Chang, and O. Painter,
Nature {\bf 472}, 69 (2011).

\bibitem{Zhou_NatPhys9_2013}
X. Zhou, F. Hocke, A. Schliesser, A. Marx, H. Huebl, R. Gross, and T. J. Kippenberg,
Nat. Phys. {\bf 9}, 179 (2013).

\bibitem{Lu_PhysRevApp10_2018}
 H. L\"u, C. Wang, L. Yang, and H. Jing,
Phys. Rev. Applied {\bf 10}, 014006 (2018).

\bibitem{Arruda_Springer219_2018}
T. J. Arruda, A. S. Martinez, F. A. Pinheiro, R. Bachelard, S. Slama, and P. W. Courteille in {\it Fano Resonances in Optics and Microwaves: Physics and Applications}, edited by E. Kamenetskii, A. Sadreev, and A. Miroshnichenko (Springer, Cham, Switzerland, 2018), pp. 445--472.


\bibitem{Arruda_PhysRevA87_2013}
T. J. Arruda, A. S. Martinez, and F. A. Pinheiro,
{Phys. Rev. A} {\bf 87}, 043841 (2013);
{\bf 92}, 023835 (2015).




\bibitem{Jiang_OptExp22_2014}
L. Sun, B. Tang, and C. Jiang,
Opt. Express {\bf 22}, 26487 (2014).

 \bibitem{Vidal_PhysRevB85_2012}
P. A. Huidobro, A. Yu. Nikitin, C. Gonz\'alez-Ballestero, L. Mart\'in-Moreno, and F. J. Garc\'ia-Vidal,
 Phys. Rev. B {\bf 85}, 155438 (2012).

\bibitem{Cuevas_JQSRT200_2017}
M. Cuevas,
J. Quant. Spectrosc. Radiat. Transf. {\bf 200}, 190 (2017);
{\bf 206}, 157 (2018);
{\bf 214}, 8 (2018).

\bibitem{Naserpour_SciRep7_2017}
M. Naserpour, C. J. Zapata-Rodr\'iguez, S. M. Vukovi\'c, H. Pashaeiad, and M. R. Beli\'c,
Sci. Rep. {\bf 7}, 12186 (2017).

\bibitem{Kleiman_JQSRT63_1999}
I. Gurwich, N. Shiloah, and M. Kleiman
J. Quant. Spectrosc. Radiat. Transf. {\bf 63}, 217 (1999).

\bibitem{Yariv_JOSAB16_1999}
Y. Xu, J. S. Vuckovic, R. K. Lee, O. J. Painter, A. Scherer, and A. Yariv,
J. Opt. Soc. Am. B {\bf 16}, 465 (1999).

\bibitem{Klimov_PhysRevA69_2004}
V. V. Klimov and M. Ducloy,
Phys. Rev. A {\bf 69}, 013812 (2004).

\bibitem{Gaponenko_JPhysChem116_2012}
D. V. Guzatov, S. V. Vaschenko, V. V. Stankevich, A. Ya. Lunevich, Y. F. Glukhov, and S. V. Gaponenko,
J. Phys. Chem. C {\bf 116}, 10723 (2012).

\bibitem{Wylie_PhysRevA30_1984}
J. M. Wylie and J. E. Sipe,
Phys. Rev. A {\bf 30}, 1185 (1984).


\bibitem{Abujetas_ACSPhot2_2015}
D. R. Abujetas, R. Paniagua-Dom\'inguez, and J. A. S\'anchez-Gil,
ACS Photonics {\bf 2}, 921 (2015).

\bibitem{Brandes_PhysRevA79_2009}
Y. N. Chen, G. Y. Chen, D. S. Chuu, and T. Brandes,
Phys. Rev. A {\bf 79}, 033815 (2009).

\bibitem{Chen_OptLett40_2015}
R. J. Li, X. Lin, S. S. Lin, X. Liu, and H. S. Chen,
Opt. Lett. {\bf 40}, 1651 (2015).

\bibitem{Falkovsky_PhysUSP51_2008}
L. A. Falkovsky,
Phys.-Usp. {\bf 51}, 887 (2008).

\bibitem{Shen_NanoLett14_2014}
W. Li, B. Chen, C. Meng, W. Fang, Y. Xiao, X. Li, Z. Hu, Y. Xu, L. Tong, H. Wang, W. Liu, J. Bao, and Y. R. Shen,
Nano Lett. {\bf 14}, 955 (2014).

\bibitem{Arruda_JOpt14_2012}
T. J. Arruda, F. A. Pinheiro, and A. S. Martinez,
J. Opt. {\bf 14}, 065101 (2012).

\bibitem{Arruda_PhysRevA94_2016}
T. J. Arruda, A. S. Martinez, and F. A. Pinheiro,
Phys. Rev. A {\bf 94}, 033825 (2016).

\bibitem{Arruda_JOSA32_2015}
T. J. Arruda, A. S. Martinez, and F. A. Pinheiro,
J. Opt. Soc. Am. A {\bf 32}, 943 (2015).

\bibitem{Zayats_OptExp21_2013}
J. Zhang and A. Zayats
Opt. Express {\bf 21}, 8426 (2013).


\bibitem{Sweeny_NatPhotonics8_2014}
T. M. Sweeney, S. G. Carter, A. S. Bracker, M. Kim, C. S. Kim, L. Yang, P. M. Vora, P. G. Brereton, E. R. Cleveland, and D. Gammon,
Nat. Photonics {\bf 8}, 442 (2014).


\bibitem{Lin_SciRep5_2015}
A. L. Feng, M. L. You, L. Tian, S. Singamaneni, M. Liu, Z. Duan, T. J. Lu, F. Xu, and M. Lin,
Sci. Rep. {\bf 5}, 7779 (2015).


\bibitem{Alu_NJPhys12_2010}
A. Al\`u, D. Rainwater, and A. Kerkhoff,
N. J. Phys. {\bf 12}, 103028 (2010).


\bibitem{Wait_CanJPhys33_1955}
J. R. Wait,
Can. J. Phys. {\bf 33}, 189 (1955).

\bibitem{Letokhov_JModOpt43_2_1996}
V. Klimov, M. Ducloy, and V. S. Letokhov,
J. Mod. Opt. {\bf 43}, 2251 (1996).

\bibitem{Dujardin_OptExp16_2008}
G. Colas des Francs, A. Bouhelier, E. Finot, J. C. Weeber, A. Dereux, C. Girard, and E. Dujardin,
Opt. Express {\bf 16}, 17654 (2008).

\bibitem{Klimov_PhysRevA62_2000}
V. V. Klimov and M. Ducloy,
Phys. Rev. A {\bf 62}, 043818 (2000).

\bibitem{Bradley_PhysRevA89_2014}
V. Karanikolas, C. A. Marocico, and A. L. Bradley,
Phys. Rev. A {\bf 89}, 063817 (2014).

\bibitem{Zakowicz_PhysRevA62_2000}
W. Zakowicz and M. Janowicz,
 Phys. Rev. A {\bf 62}, 013820 (2000).

\end{thebibliography}
\end{document}